%% file: article.tex
\pdfoutput=1
\documentclass[11pt,onecolumn]{IEEEtran}

\usepackage[latin1]{inputenc}
\usepackage{graphicx}
\usepackage[gray]{xcolor}
\usepackage{amsmath,amsfonts,amssymb}
\usepackage{bbm}
\usepackage{cite}

\usepackage{hyperref}
\usepackage[T1]{fontenc}
\usepackage{lmodern}
\newtheorem{theorem}{Theorem}
\newtheorem{definition}{Definition}
\newtheorem{property}{Property}
\newtheorem{lemma}{Lemma}
\newcommand{\erfc}[1]{\operatorname{erfc}\left(#1\right)}

\title{Robustness maximization of parallel multichannel systems}

\author{\thanks{The reserach leading to these results has received
    partial funding from the European Community's Seventh Framework
    Programme FP7/2007-2013 under grand aggreement
    n\textsuperscript{o}\,213311 also referred to as OMEGA.}Jean-Yves
  \textsc{Baudais}\thanks{J.-Y. Baudais is with the National Center
    for Scientific Research (CNRS) in the Institute of Electronics and
    Telecommunications of Rennes (IETR), UMR 6164, F-35708 Rennes,
    France (email: jean-yves.baudais@insa-rennes.fr).}, Fahad
  \textsc{Syed Muhammad}, and Jean-Fran\c cois \textsc{H\'elard},
  \IEEEmembership{Senior Member, IEEE}\thanks{F. Syed Muhammad and
    J.-F.  H\'elard are with the \emph{Universit\'e europ\'eenne de
      Bretagne}, INSA, IETR, UMR 6164, F-35708 Rennes, France (email:
    \{fahad.syed-muhammad;jean-francois.helard\}@insa-rennes.fr).}}

\begin{document}

\maketitle
\IEEEpeerreviewmaketitle

\input texte.tex
\input appendix.tex

\bibliographystyle{IEEEtran}
%\IEEEtriggeratref{42}
\input article.bbl

%\bibliography{macro,bibdoc,bibperso}

\end{document}

%% file: texte.tex
\begin{abstract}
  Bit error rate (BER) minimization and SNR-gap maximization, two
  robustness optimization problems, are solved, under average power
  and bit-rate constraints, according to the waterfilling policy.
  Under peak-power constraint the solutions differ and this paper
  gives bit-loading solutions of both robustness optimization problems
  over independent parallel channels.  The study is based on
  analytical approach with generalized Lagrangian relaxation tool and
  on greedy-type algorithm approach. Tight BER expressions are used
  for square and rectangular quadrature amplitude modulations. Integer
  bit solution of analytical continuous bit-rates is performed with a
  new generalized secant method.  The asymptotic convergence of both
  robustness optimizations is proved for both analytical and
  algorithmic approaches. We also prove that, in conventional margin
  maximization problem, the equivalence between SNR-gap maximization
  and power minimization does not hold with peak-power limitation.
  Based on a defined dissimilarity measure, bit-loading solutions are
  compared over power line communication channel for multicarrier
  systems.  Simulation results confirm the asymptotic convergence of
  both allocation policies. In non asymptotic regime the allocation
  policies can be interchanged depending on the robustness measure and
  the operating point of the communication system.  The low
  computational effort of the suboptimal solution based on analytical
  approach leads to a good trade-off between performance and
  complexity.
\end{abstract}

\section{Introduction}

\IEEEPARstart{I}{n transmitter} design, a problem often encountered is
resource allocation among multiple independent parallel channels.  The
resource can be the power, the bits or the data and the number of
channels. The allocation policies are performed under constraints and
assumptions, and the independent parallel channels can be encountered
in multitone transmission or multiantenna communications.

Independent parallel channels result from orthogonal design applied in
time, frequency or spatial domains \cite{AKAN98}.  They can either be
obtained naturally or in a situation where the transmit and receive
strategies are to orthogonalize multiple waveforms.  Orthogonal
frequency-division multiplexing (OFDM) and digital multitone (DMT) are
two successful commercial applications for wireless and wireline
communications with orthogonality in the frequency domain. In
multiantenna communications, the parallel channels are made from the
singular vectors of the multiple-input multiple-output (MIMO) channel
\cite{TELA95}.  This MIMO concept and the resulting orthogonal design
can be applied in many communication scenarios when there are multiple
transmit and receive dimensions \cite{PALO06b}.

To perform allocation, mathematical relations between various
resources are needed and the first one is the channel capacity. This
capacity of $n$ independent parallel Gaussian channels is the
well-known sum of the capacities of each channel
\begin{align}\label{eq_capai}
  \mathcal{C}=\sum_{i=1}^n\mathcal{C}_i=\sum_{i=1}^n\log_2(1+\mathsf{snr}_i)\,.
\end{align}
This relation, which holds for memoryless channels, links the supremum
bit-rate $\mathcal{C}_i$, here expressed in bit per two dimensions, to
the signal to noise ratio, $\mathsf{snr}_i$, experienced by each
channel or subchannel $i$. Any reliable and implementable system must
transmit at a bit-rate $r_i$ below capacity $\mathcal{C}_i$ over each
subchannel and then the margin, or SNR-gap, $\gamma_i$ is introduced
to analyze such systems \cite{CIOF91, FORN91}
\begin{align}\label{eq_giinit}
  \gamma_i=\frac{2^{\mathcal{C}_i}-1}{2^{r_i}-1}\,.
\end{align}
This SNR-gap is a convenient mechanism for analyzing systems that
transmit below capacity, and
\begin{align}\label{eq_riinit}
  r_i=\log_2\Big(1+\frac{\mathsf{snr}_i}{\gamma_i}\Big),
\end{align}
with $r_i$ the bit-rate in bits per two dimensional-symbol (bits per
second per subchannel) which is also the number of bits per
constellation symbol.

Resource allocation is performed using loading algorithms and diverse
criteria can be invoked to decide which portion of the available
resource is allocated to each of the subchannels. From information
theory point of view, the criterion is the mutual information and the
optimal allocation under average power constraint was first devised in
\cite{SHAN49} for Gaussian inputs and later for non-Gaussian inputs
\cite{LOZA06}. Since the performance measure is the mutual information
in these cases, the SNR-gap in \eqref{eq_riinit} is
$\{\gamma_i\}_{i=1}^n=1$. In other cases, $\{\gamma_i\}_{i=1}^n>1$ and
\eqref{eq_riinit}, as enhanced relationships, leads to many optimal
and suboptimal allocation policies. In fact, resource allocation is a
constraint optimization problem and generally two cases are of
practical interest: rate maximization (RM) and margin maximization
(MM), where the objective is the maximization of the data or bit-rate,
and the maximization of the system margin (or power minimization in
practice), respectively \cite{PAPA08}. The MM problem gathers all non
RM problems including power minimization, margin maximization (in its
strict sense) and other measures such as error probabilities or
goodput\footnote{In this paper MM abbreviation is related to the
  general family of non RM problems and not only to the margin
  maximization problem in its strict sense. The expanded form is
  reserved for the margin maximization in its strict sense.}.  Among
all the resource allocation strategies, equivalence or duality can be
found defining family of approaches and using unified processes
\cite{CHUN01, FASA04, PALO05, KIM06}.  The loading algorithms are also
split in two families.  The first is based on greedy-type approach to
iteratively distribute the discrete resources \cite{HUGH87}, and the
second uses Lagrangian relaxation to solve continuous resource
adaptation \cite{KRON00}. Both approaches have been compared in terms
of performance and complexity \cite{CHOI00, KRON00, CHUN01, UTHA05}.
All these adaptive resource allocations are possible when channel
state information (CSI) is known at both transmitter and receiver
sides.  This CSI can be perfect or imperfect, and full or partial. The
effects of channel estimation error and feedback delay on the
performance of adaptive modulated systems can also be considered in
the allocation process \cite{GOLD97a, YE06, ERMO08}.

In this paper we shall focus henceforth on MM problems with perfect
and full CSI consideration, and under peak-power and bit-rate
constraints. It is assumed that the channel estimation is perfect, and
feedback CSI delay and overhead are negligible. Peak-power constraint
results from power mask limitation and has been taken into account in
resource allocation problem \cite{CHOI00, BACC02, FASA03, KHOJ04}
instead of the conventional average power constraint, or sum power
constraint, which is historically the first considered constraint
\cite{SHAN49}. Bit-rate constraint comes from communication
applications or service requirements, where different flows can exist
but one of them is chosen at the beginning of the communication. In
this configuration, the remaining parameter to optimize is then the
SNR-gap $\gamma_i$ which is also related to the error probability of
the communication system.

Two similar problems of MM have the same objective that is to maximize
the system robustness.  What we call robustness in this paper is the
capability of a system to maintain acceptable performance with
unforeseen disturbances. 

The first measure of robustness is the SNR-gap, or system margin, and
its maximization ensures protection to unforeseen channel impairments
or noise. The system margin maximization is the maximization of the
minimal SNR-gap $\gamma_i$ in \eqref{eq_riinit} over the $n$
subchannels. In that case the conventional equivalence between margin
maximization and power minimization in MM problems is not generally
true. In this paper we show that this equivalence can nevertheless be
obtained in particular configurations.

The second robustness measure is the bit error rate (BER) and its
minimization can reduce the packet error rate and the data
retransmission.  In transmitter design, the BER minimization can be
realized using uniform bit-loading and adaptive precoding
\cite{DING03, PALO03}. Analytical studies have been performed with
peak-BER or average BER (computed as arithmetic mean) approaches
\cite{GOLD97a, ERMO08}. With average BER computed as weighted
arithmetic mean, the resource allocation has been performed using
greedy-type algorithm \cite{WYGL05}. The first main contribution in
this paper is the analytical solution of the resource allocation
problem in the case of weighted arithmetic mean BER minimization.

To perform the analytical study based on generalized Lagrangian
relaxation tool, we develop a new method for finding roots of
functions. This method generalizes the secant method to better fit the
function-depending weight and to speed up the search of the roots.
Both robustness polices are compared with a new measure that evaluates
the difference in the bit distribution between two allocations. We
also prove that both robustness policies provide the same bit
distribution in asymptotic regime and this is the second main
contribution in this paper.  The proof is given in the case of
unconstrained modulations (i.e.\ continuous bit-rates and analytical
solution) and also for QAM constellations and greedy-type algorithms.
The convergence is exemplified by simulation in multicarrier PLC
(power line communications) systems.

The organization of the paper is as follows. In Section~\ref{sec_pb},
the quantities to be used throughout the paper are introduced and the
robustness optimization problem is formulated in a general way for
both system margin maximization and BER minimization. The cases of
equivalence between margin maximization and power minimization are
worked out.  Section~\ref{sec_inter} is an interlude, presenting the
considered expressions of accurate BER, a new measure of allocation
differences and a new search method of root function. The solution of
formulated problems are given in Section~\ref{sec_iter} in the form of
an optimum allocation policy based on greedy-type algorithms. The
conditions of equivalence of both margin maximization and BER
minimization are given in this section.  Section~\ref{sec_anal}
presents the analytical solution and both greedy-type and analytical
methods are compared in Section~\ref{sec_cmp}.  In turn, Section
\ref{sec_sim} exemplifies the application of robustness optimization
to multicarrier PLC systems.  Finally, the paper concludes in Section
\ref{sec_conc} with the proofs of several results relegated to the
appendices.

Notation. The bit-rates $\{r_i\}_{i=1}^n$ are defined as a number of
bits per two dimensions and they also could only be a number of bits
(undertone per constellation). Without confusing, the unit of the
variable $r_i$ are not all the time fully expressed.

\section{Problem formulation}\label{sec_pb}

Consider $n$ parallel subchannels. On the $i$th subchannel, the
input-output relationship is
\begin{align}
  Y_i=h_iS_i+W_i\ ,
\end{align}
with $S_i$ the transmitted symbol, $Y_i$ the received one, and $h_i$
the complex scalar channel gain. The complex Gaussian noise $W_i$ is a
proper complex random variable with zero-mean and variance equal to
$\sigma_{W_i}^2$.

The conventional average power constraint is
\begin{align}
  \frac{1}{n}\sum_{i=1}^nE[|S_i|^2]\leq P\ ,
\end{align}
whereas the peak-power constraint, or power spectrum density
constraint, considered in this paper is
\begin{align}
  \forall i\in[1,n]\quad E[|S_i|^2]\leq P\,.
\end{align}

It is convenient to use normalized unit-power symbol $\{X_i\}_{i=1}^n$
such that
\begin{align}
  S_i=\sqrt{p_iP}X_i\,,
\end{align}
which leads to the peak-power constraint
\begin{align}
  \forall i\in[1,n]\quad p_i\leq1\,.
\end{align}

It is also convenient to introduce two other variables. The first one
is the conventional SNR
\begin{align}
  \mathsf{snr}_i=|h_i|^2p_i\frac{P}{\sigma_{W_i}^2}
\end{align}
and the second is called \emph{power spectrum density noise ratio}
(PSDNR)
\begin{align}
  \mathsf{psdnr}=\frac{1}{n}\sum_{i=1}^n|h_i|^2\frac{P}{\sigma_{W_i}^2}\,,
\end{align}
which is the mean signal to noise ratio over the $n$ subchannels if
and only if $\forall i$ $p_i=1$.  This PSDNR is the ratio between the
power mask at receiver side (the transmitted power mask through the
channel) and the power spectrum density of the noise.  The system
performance will be given according to this parameter to point out the
ability of a system to exploit the available power under peak-power
constraint.

Using the previous notations, \eqref{eq_riinit} becomes
\begin{align}\label{eq_ri}
  r_i=&\log_2\Big(1+\frac{|h_i|^2p_iP}{\gamma_i\sigma_{W_i}^2}\Big)\,.
\end{align}
With $p_i/\gamma_i=1$ for all $i$, $r_i$ is the subchannel capacity
under power constraint $P$.  With unconstrained modulations, $r_i$ is
defined in $\mathbb{R}$, but constrained modulations are used in
practice and $r_i$ takes a finite number of nonnegative values. Non
integer number of bits per symbol can also be used with fractional bit
constellations~\cite{FORN84, CIOF07}. In this paper, modulations
defined by discrete points are used with integer number of bits per
symbol. Typically,
$r_i\in\{0{,}\beta{,}2\beta{,}\cdots{,}r_\text{max}\}$, where $\beta$
is the granularity in bits and $r_\text{max}$ is the number of bits in
the richest available constellation. When all QAM constellations are
used $\beta=1$. The peak-power and bit-rate constraints are then
\begin{align}
  \forall i\ p_i\leq1,\
  \displaystyle\sum_{i=1}^nr_i=R,\
  \forall i\ r_i\in\{0{,}\beta{,}2\beta{,}\cdots{,}r_\text{max}\}\,.
\end{align}
Obviously, the exploitation of available power leads to $\forall i\
p_i=1$ and the constraint is simplified
\begin{align}\label{eq_const}
  \displaystyle\sum_{i=1}^nr_i=R,\quad \forall i\
  r_i\in\{0{,}\beta{,}2\beta{,}\cdots{,}r_\text{max}\}\,.
\end{align}
With peak-power and bit-rate constraints, the allocation strategy is
then to use all available power and to optimize the robustness.

The problem we pose is to determine the optimal bit-rate allocation
$\{r_i^*\}_{i=1}^n$ that maximizes a robustness measure, or inversely
minimizes a frailness measure, with given SNR of subchannels while
satisfying \eqref{eq_const}. In its general form, this problem can be
written as
\begin{align}\label{eq_pbgen}
  [r_1^*,\cdots,r_n^*]=\arg\min_{\eqref{eq_const}}\phi\Big(\{r_i\}_{i=1}^n\Big),
\end{align}
where $\phi(\cdot)$ is the frailness measure. In this paper, this
measure is given by the SNR-gap or the BER. In addition to the
bit-rate allocation, the receiver is presumed to have knowledge of the
magnitude and phase of the channel gain $\{h_i\}_{i=1}^n$, whereas the
transmitter needs only to know the magnitude $\{|h_i|\}_{i=1}^n$. The
objective is to find the data vector $[r_1^*,\cdots,r_n^*]$ which is
the final relevant information for the transmitter. The allocation can
then be computed on the receiver side to reduce the feedback data-rate
from $n$ real numbers to $n$ finite integer numbers.  Furthermore, the
integer nature of the data-rates allows a full CSI at the transmitter
which is not possible with real numbers.

\subsection{System margin maximization}

The SNR-gap $\gamma_i$ of the subchannel $i$ is \eqref{eq_riinit}
\begin{align}
  \gamma_i=\frac{\mathsf{snr}_i}{2^{r_i}-1}\,.
\end{align}
With reliable communications, $\gamma_i$ is higher than 1 for all
subchannels.  Let the system margin, or system SNR-gap, be the minimal
value of the SNR-gap in each subchannel
\begin{align}\label{eq_gapdef}
  \gamma=\min_i\gamma_i\,.
\end{align}
Let $\gamma_\text{init}$ be the initial system margin of one
communication system ensuring a given QoS. Let $\gamma$ be the
optimized system margin of this system. Then, the system margin
improvement ensure system protection in unforeseen channel impairment
or noise, e.g.\ impulse noise: bit-rate and system performance targets
are always reached for an unforeseen SNR reduction of
$\gamma/\gamma_\text{init}$ over all subchannels.  This robustness
optimization does not depend on constellation and channel coding
types. The system margin $\gamma$ is defined and optimized without
knowledge of used constellations and coding, and the proposed
robustness optimization works for any coding and modulation scheme.

The objective is the maximization of the system margin which is
equivalent to the minimization of $\gamma^{-1}$. We note
$\gamma_i(r_i)$ the function that associates $r_i$ to $\gamma_i$.  The
function $\phi(\cdot)$ in \eqref{eq_pbgen} is then given by
\begin{align}\label{eq_phigap}
  \phi\Big(\{r_i\}_{i=1}^n\Big)=\max_i\frac{1}{\gamma_i(r_i)}
\end{align}
and
\begin{align}\label{eq_pbgap}
  [r_1^*{,}\cdots{,}r_n^*]=\arg\min_\eqref{eq_const}\max_i\gamma_i^{-1}\,.
\end{align}
This problem is the converse problem of bit-rate maximization under
peak-power and SNR-gap constraints.  The solution of the bit-rate
maximization problem is obvious under the said constraints and given
by
\begin{align}
  \forall i\quad r_i^*=\beta\left\lfloor\frac{1}{\beta}\log_2\left(1+\frac{\mathsf{snr}_i}{\gamma_i}\right)\right\rfloor.
\end{align}

Following the conventional SNR-gap approximation \cite{CIOF91}, symbol
error rate (SER) expression of QAM depending on SNR-gap is
constellation size-independent with
\begin{align}
  \forall r_i\quad\mathsf{ser}_i(r_i)=2\erfc{\sqrt{\frac{3}{2}\gamma_i}}\,,
\end{align}
where the complementary error function is usually defined as
\begin{align}
  \erfc{x}=\frac{2}{\sqrt{\pi}}\int_x^\infty e^{-t^2}\operatorname{d}\!t\,.
\end{align}
The system margin maximization is then equivalent to the peak-SER
minimization in high SNR regime.  Note that with \eqref{eq_gapdef},
the system margin maximization can also be called a trough-SNR-gap
maximization and it strongly related to the peak-power minimization.
Whereas the bit-loading solution is the same for power minimization
and margin maximization, with sum-margin or sum-power constraints
instead of peak constraints, the following lemma gives sufficient
conditions for equivalence in the case of peak constraints.

\begin{lemma}\label{theo_eqmarpow}
  The bit allocation, that maximizes the system margin under
  peak-power constraint $\{p_i^\text{margin}\}_{i=1}^n$, minimizes the
  peak-power under SNR-gap constraint
  $\{\gamma_i^\text{power}\}_{i=1}^n$ if
  $p_i^\text{margin}\gamma_i^\text{power}=\alpha$ for all $i$.
\end{lemma}

\begin{IEEEproof}
  It is straightforward using \eqref{eq_ri} and \eqref{eq_pbgap}. Both
  problems have the same expression and therefore the same solution.
\end{IEEEproof}

This lemma provides a sufficient but not necessary condition for the
equivalence of solutions, and it says that if the power and the
SNR-gap constraints have proportional distributions for margin
maximization and peak-power minimization problems, respectively, then
both problems have the same optimal bit-rate allocation. This lemma
also shows that both problems don't have the same solution in a
general case. A particular case is the uniform distribution of
$\{p_i\}_{i=1}^n$ which is the case analyzed in this paper.

\subsection{BER minimization}

In communication systems, the error rate of the transmitted bits is a
conventional robustness measure. By definition, the BER is the ratio
between the number of wrong bits and the number of transmitted bits.
With a multidimensional system, there exists several BER expressions
\cite{GOLD97a, WYGL05}.  Let the BER evaluated over the transmission
of $m$ multidimensional symbols\footnote{We suppose that $m$ is high
  enough to respect the ergodic condition and to make possible use of
  error probability.}.  In our case, the multidimensional symbols are
the symbols sent over $n$ subchannels. Let $e_i$ the number of
erroneous bits received over subchannel $i$ during the transmission.
The BER is then given as
\begin{align}
  \mathsf{ber}=\frac{\sum\limits_{i=1}^ne_i}{m\sum\limits_{i=1}^nr_i}=\frac{\sum\limits_{i=1}^nr_i\frac{e_i}{mr_i}}{\sum\limits_{i=1}^nr_i}\,.
\end{align}
The BER over subchannel $i$ is $e_i/mr_i$ and the BER of $n$
subchannels is then
\begin{align}\label{eq_ber}
  \mathsf{ber}\Big(\{r_i\}_{i=1}^n\Big)=
  \frac{\sum\limits_{i=1}^nr_i\mathsf{ber}_i(r_i)}{R}\,.
\end{align}
The BER of multiple variable bit-rate $r_i$ is then not the arithmetic
mean of BER but is the weighted mean BER. Weighted mean BER and
arithmetic mean BER are equal if $\forall i,j$ $r_i=r_j$ or if
$\mathsf{ber}_i=0$ for all $i$. As there exists $\mathsf{ber}_i\neq0$
then weighted mean BER and arithmetic mean BER are equal if and only
if $\forall i,j$ $r_i=r_j$. Note that if the number $m$ of transmitted
multidimensional symbols depend on the subchannel $i$, \eqref{eq_ber}
does not hold anymore.

The function $\phi(\cdot)$ in \eqref{eq_pbgen} is then given by
\begin{align}\label{eq_phiber}
  \phi\Big(\{r_i\}_{i=1}^n\Big)=\frac{1}{R}\sum\limits_{i=1}^nr_i
  \mathsf{ber}_i(r_i)
\end{align}
and
\begin{align}\label{eq_pbber}
  [r_1^*{,}\cdots{,}r_n^*]=\arg\min_{\eqref{eq_const}}\mathsf{ber}\Big(
  \{r_i\}_{i=1}^n\Big)\,.
\end{align}

To simplify notations, let $\mathsf{ber}(R)$ the BER of the system. In
high SNR regime with Gray mapping,
$r_i\mathsf{ber}_i(r_i)=\mathsf{ser}_i(r_i)$ and then weighted mean
BER can be approximated by arithmetic mean SER divided by the number
of transmitted bits.

Contrary to system margin maximization, the BER minimization needs the
knowledge of constellation and coding schemes and it is based on valid
expressions of BER functions. In this paper, the used constellations
are QAM and the optimization is performed without channel coding
scheme.  When dealing with practical coded systems, the ultimate
measure is the coded BER and not the uncoded BER.  However, the coded
BER is strongly related to the uncoded BER. It is then generally
sufficient to focus on the uncoded BER when optimizing the uncoded
part of a communication system \cite{PALO04}.

\section{Interludes}\label{sec_inter}

Before solving the optimization problem, the BER approximation of QAM
is presented. This approximation plays a chief role in BER
minimization, a good approximation is therefore needed.  Since this
paper deals with bit-rate allocation, a measure of difference in the
bit-rate distribution is proposed and presented in this section. This
section also presents a new research method of root function. This
method generalizes the secant method and converges faster than the
secant one.

\subsection{BER approximation}
                                                    
Conventionally, the BER approximation of square QAM has been performed
by either calculating the symbol error probability or by simply
estimating it using lower and upper bounds~\cite{PROA95}. This
conventional approximation tends to deviate from its exact values when
SNR is low and it cannot be applied for rectangular QAM. Exact and
general closed-form expressions are developed in~\cite{CHO02} for
arbitrary one and two-dimensional amplitude modulation schemes. 

An approximate BER expression for QAM can be obtained by neglecting
the higher order terms in the exact closed-form expression
\cite{CHO02}
\begin{align}\label{eq_appr}
  \mathsf{ber}_i\simeq\frac{1}{r_i}\Big(2-\frac{1}{I_i}-\frac{1}{J_i}\Big)\erfc{\sqrt{\frac{3}{I_i^2+J_i^2-2}\,\mathsf{snr}_i}}
\end{align}
with $I_i=2^{\lfloor r_i/2\rfloor}$, $J_i=2^{\lceil r_i/2\rceil}$ and
$r_i=\log_2(I_i\cdot J_i)$. By symmetry, $I_i$ and $J_i$ can be
inverted. The BER can also be expressed using the SNR-gap $\gamma_i$.
Using \eqref{eq_riinit} and \eqref{eq_appr}, the BER is written as
\begin{align}
  \mathsf{ber}_i\simeq\frac{1}{r_i}\Big(2-\frac{1}{I_i}-\frac{1}{J_i}\Big)\erfc{\sqrt{\frac{3(I_iJ_i-1)}{I_i^2+J_i^2-2}\,\gamma_i}}
\end{align}

\begin{figure}
  \centering
  \includegraphics[width=\linewidth]{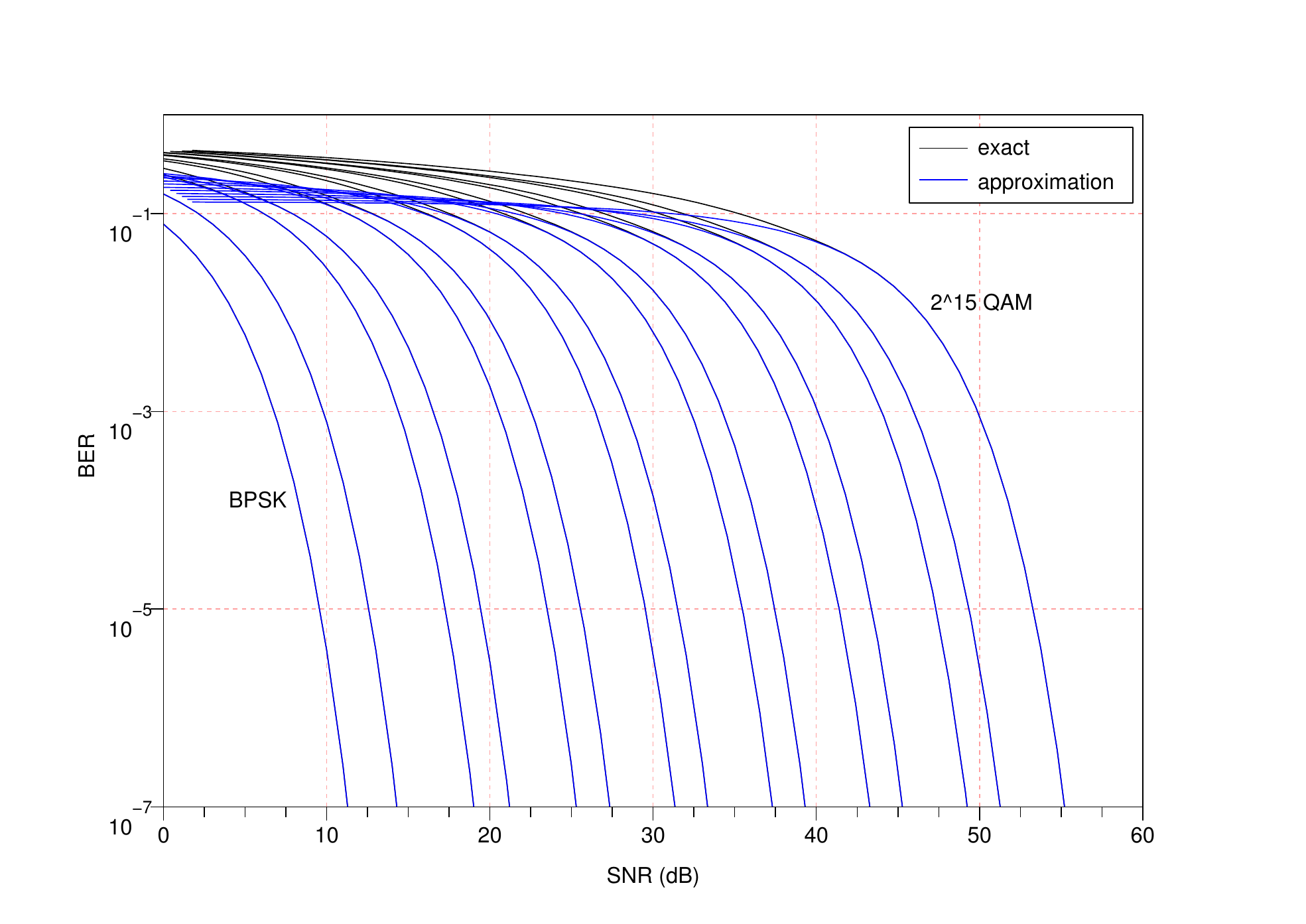}
  \caption{Exact BER curves and approximations \eqref{eq_appr}.}
  \label{fig_ber}
\end{figure}

These two approximations allow extension of the $\mathsf{ber}_i(r_i)$
function from $\mathbb{N}$ to $\mathbb{R}_+$ which is useful for
analytical studies. Fig.~\ref{fig_ber} gives the theoretical
BER-curves and the
approximated ones from the binary phase shift keying (BPSK) to the
32768-QAM. For BER lower than $5\cdot10^{-2}$ the relative error is
lower than 1~\% for all modulations.

\subsection{Dissimilar allocation measure}

To measure the difference in the bit distribution between different
allocation strategies, we need to evaluate the \emph{dissimilarity}.
This dissimilarity measure must verify the following properties: 1) if
two allocations lead to the same bit distribution then the measure of
dissimilarity must be null, whereas 2) if two allocations lead to two
completely different bit distributions in loaded subchannels, then the
measure of dissimilarity must be equal to one, and 3) the measure is
symmetric, i.e. the dissimilarity between allocation $X$ and
allocation $Y$ must be the same as the dissimilarity between
allocation $Y$ and allocation $X$. We choose that the empty
subchannels do not impact the measure.

\begin{definition}\label{def_diss}
  The dissimilarity measure between allocation $X$ and allocation $Y$
  is
  \[
  \mu(X,Y)=\frac{\displaystyle\sum_{i=1}^n\delta\Big(r_i(X)-r_i(Y)\Big)}{\displaystyle\max_{j\in\{X,Y\}}\sum_{i=1}^n\delta\Big(r_i(j)\Big)}
  \]
  where $\delta(x)=1$ if $x\neq0$ else $\delta(x)=0$.
\end{definition}

This dissimilarity has the following properties.
\begin{property}
  $\mu(X,Y)=0$ iff $\forall i$, $r_i(X)=r_i(Y)$.
\end{property}
\begin{property}
  $\mu(X,Y)=1$ iff $\forall i$ then $r_i(X)\neq r_i(Y)$ or
  $r_i(X)=r_i(Y)=0$.
\end{property}
\begin{property}
  $\mu(X,Y)=\mu(Y,X)$.
\end{property}
\begin{property}
  If $\mu(X,Y)=0$, then for all allocation $Z$, $\mu(X,Z)=\mu(Y,Z)$.
\end{property}

All these properties are direct consequences of
Definition~\ref{def_diss}. For a null dissimilarity, $\mu(X,Y)=0$, all
the subchannels transmit the same number of bits, i.e. $\forall i$
$r_i(X)=r_i(Y)$. For a full dissimilarity, $\mu(X,Y)=1$, all the non
empty subchannels of both allocations $X$ and $Y$ transmit a different
number of bits, i.e.\ $\forall i$ such as $r_i(X)\neq0$ and
$r_i(Y)\neq0$ then $r_i(X)\neq r_i(Y)$.  It is obvious that the
measure is symmetric $\mu(X,Y)=\mu(Y,X)$. If two allocations have a
null dissimilarity $\mu(X,Y)=0$, then they are identical and for any
allocation $Z$ $\mu(X,Z)=\mu(Y,Z)$. The converse of this last property
is not true. Note that the dissimilarity is not defined for two empty
allocations.

For example, let $n=4$ and $[r_1(X),\cdots,r_4(X)]=[4\;3\;3\;0]$. If
$[r_1(Y),\cdots,r_4(Y)]=[3\;2\;2\;2]$ or
$[r_1(Y),\cdots,r_4(Y)]=[5\;5\;0\;0]$ then $\mu(X,Y)=1$. If
$[r_1(Y),\cdots,r_4(Y)]=[4\;3\;2\;1]$ then $\mu(X,Y)=\frac{1}{2}$. The
measure $\mu(X,Y)$ is null if and only if
$[r_1(Y),\cdots,r_4(Y)]=[4\;3\;3\;0]$.

\subsection{Generalized secant method}

There are many numerical methods for finding roots of function. We
propose a new method, called generalized secant method that is based
on secant method. This new method better fits the function-depending
weight than secant method to speed the convergence. Before explaining
this new method, a brief overview of secant method is given.

In our case, the objective function $f(x)$ is monotonous, non
differentiable and computable over $x\in[x_1,x_2]$ with
$\frac{f(x_1)}{|f(x_1)|}=-\frac{f(x_2)}{|f(x_2)|}$.  The secant method
is as follows for an increasing function $f(x)$.
\begin{enumerate}
\item $i=0$, $y_0=f(x_1)$;
\item $x_0=\frac{x_2f(x_1)-x_1f(x_2)}{f(x_1)-f(x_2)}$, $y_{i+1}=f(x_0)$;
\item If $|y_{i+1}-y_i|\leq\epsilon$ then $x_0$ is the root of $f(x)$,
  else $\left\{\begin{array}{l}y_{i+1}<0\text{ then
      }x_1=x_0\\y_{i+1}>0\text{ then }x_2=x_0\end{array}\right\}$,
  $i\to i+1$ and go to step~2.
\end{enumerate}
The objective of the secant method is to approximate $f(x)$ by a
linear function $g_i(x)=a_ix+b_i$ at each iteration $i$, with
$g_i(x_1)=f(x_1)$ and $g_i(x_2)=f(x_2)$, and to set $x_0$ as the root
of $g_i(x)$. The search for the root of $f(x)$ is completed when the
desired precision $\epsilon$ is reached. The precision is given for
$y_i$ but it can also be given for $x_i$.

As the function $f(x)$ is computable, it can be plotted and an \emph{a
  posteriori} simple algebraic or elementary transcendental invertible
function over $[x_1,x_2]$ can be used to better fit the function
$f(x)$. An \emph{a posteriori} information is then used to improve the
search for the root. The function $f(x)$ is iteratively approximated
by $a_ih(x)+b_i$ instead of $a_ix+b_i$, where $h(x)$ is the invertible
function.  This method is then given as follows for an increasing
function $f(x)$.
\begin{enumerate}
\item $i=0$, $y_0=f(x_1)$;
\item $x_0=h^{-1}\Big(\frac{x_2f(x_1)-x_1f(x_2)}{f(x_1)-f(x_2)}\Big)$,
  $y_{i+1}=f(x_0)$;
\item If $|y_{i+1}-y_i|\leq\epsilon$ then $x_0$ is the root of $f(x)$,
  else $\left\{\begin{array}{l}y_{i+1}<0\text{ then
      }x_1=x_0\\y_{i+1}>0\text{ then }x_2=x_0\end{array}\right\}$,
  $i\to i+1$ and go to step~2.
\end{enumerate}
Compared to secant method, only step 2) differs from generalized
secant method where the computation of $x_0$ is performed taking into
account the approximated shape $h(x)$ of the function~$f(x)$.

This generalized secant method is used in Section~\ref{sec_anal} to
find the root of the Lagrangian and is compared to the conventional
secant method. In our case, $f(x)$ is the sum of logarithmic functions
and the function $h(x)$ is then the logarithmic one.

\section{Optimal greedy-type allocations}\label{sec_iter}

The general problem is to find the optimal allocation
$[r_1^*,\cdots,r_n^*]$ that minimizes $\phi(\cdot)$, the inverse
robustness measure, or frailness. This is a combinatorial optimization
problem or integer programming problem.  The core idea in this
iterative allocation is that a sequential approach can lead to a
globally optimum discrete loading. Greedy-type methods then converge
to optimal solution under conditions. Convexity is not required for
the convergence of the algorithm and monotonicity is
sufficient~\cite{FOX66}. This monotonicity ensures that the removal or
addition of $\beta$ bits at each iteration converges to the optimal
solution.  In this paper the used functions $\phi(\cdot)$ are
monotonic increasing functions.

In its general form and when the objective function $\phi(\cdot)$ is
not only a weighted sum function, the iterative algorithm is
\begin{enumerate}
\item Start with allocation $[r_1^{(0)},\cdots,r_n^{(0)}]=0$,
\item $k=0$,
\item Allocate one more bit to the subchannel $j$ for which
  \begin{align}\label{eq_fox}
    \phi\Big(\{r_i^{(k+1)}\}_{i=1}^n\Big)
  \end{align}
  is minimal, with $r_j^{(k+1)}=r_j^{(k)}+\beta$ and $\forall i\neq j$
  $r_i^{(k+1)}=r_i^{(k)}$,
\item If $\sum_ir_i^{(k+1)}=R$, terminate; otherwise $k\to k+1$ and go
  to step 3.
\end{enumerate}

The obtained allocation is then optimal~\cite{FOX66} and solves
\eqref{eq_pbgen}. This algorithm needs $R/\beta$ iterations and its
complexity is $\mathcal{O}(nR)$. The target bit-rate $R$ is supposed
to be feasible, i.e.\ $R$ is a multiple of $\beta$. Note that an
equivalent formulation can be given starting with
$r_i^{(0)}=r_\text{max}$ for all $i$ and using bit-removal instead of
bit-addition with maximization instead of minimization. For very high
bit-rate, higher than $\frac{n}{2}r_\text{max}$, the number of
iterations with bit-removal is lower than those obtained with
bit-addition. This is the opposite with bit-rate lower than
$\frac{n}{2}r_\text{max}$.

Iterative allocations have been firstly applied to bit-rate
maximization under power constraint~\cite{HUGH87}. Many works have
been devoted to complexity reduction of greedy-type algorithms, see
for example~\cite{CHOW95, CAMP98a, KRON00, PAPA08} and references
therein. In this section, only greedy-type algorithms are presented in
order to compare the analytical allocation to the optimal iterative
one. Note that analytical solution can also be used as an input of
greedy-type algorithm to initialize it and reduce its complexity.

\subsection{System margin maximization}

The system margin, or system SNR-gap, maximization under bit-rate and
peak-power constraints is the converse problem of the bit-rate
maximization under SNR-gap and peak-power constraints. This converse
problem has been solved, e.g., in~\cite{BACC02}. To comply with the
general problem formulation, the inverse system margin minimization is
presented instead of the system margin maximization.

\begin{lemma}\label{theo_itergap} 
  Under bit-rate and peak-power constraints, the greedy-type
  allocation that minimizes the inverse system margin $\gamma^{-1}$
  \eqref{eq_gapdef} allocates sequentially $\beta$ bits to the
  subchannel $i$ bearing $r_i$ bits and for which
  \[\frac{2^{r_i+\beta}-1}{\mathsf{snr}_i}\]
  is minimum.
\end{lemma}
\begin{IEEEproof}
  It is straightforward using \eqref{eq_phigap} and \eqref{eq_fox}.
  See Appendix~\ref{ann_itergap} for an original proof.
\end{IEEEproof}

The main advantage of system margin maximization is that the optimal
allocation can be reached independently of the SNR regime. Allocation
is always possible even for very low SNR but it can lead to unreliable
communication with SNR-gap lower than 1. Lemma~\ref{theo_itergap} is
given with unbounded modulation orders, i.e.\ $r_\text{max}=\infty$
and $\forall i$ $r_i\in\beta\mathbb{N}$. With full
constraints~\eqref{eq_const}, the subchannels that reach
$r_\text{max}$ are simply removed from the iterative process.

\subsection{BER minimization}

The system BER minimization under bit-rate and peak-power constraints
is the converse problem of bit-rate maximization under peak-power and
BER constraints. This converse problem has been solved, e.g., in
\cite{WYGL05}. Using \eqref{eq_fox} and \eqref{eq_phiber}, the
solution of BER minimization is straightforward and the corresponding
greedy-type algorithm is also known as Levin-Campello algorithm
\cite{LOZA06, LEVI01, CAMP99}. The main drawback of this solution is
that it requires good approximated BER expressions even in low SNR
regime. This constraint can be relaxed and the following lemma gives
the optimal greedy-type allocation for the BER minimization.

\begin{lemma}\label{theo_iter}
  In high SNR regime and under bit-rate and peak-power constraints,
  the greedy-type allocation that minimizes the BER minimizes
  $(r_i+\beta)\mathsf{ber}_i(r_i+\beta)$ at each step.
\end{lemma}
\begin{IEEEproof}
  See Appendix~\ref{ann_iter}.
\end{IEEEproof}

Lemma~\ref{theo_iter} states how to allocate bits without mean BER
computation at each step. It is given without modulation order
limitation.  Like system margin maximization solution, the bounded
modulation order is simply taken into account using $r_\text{max}$ and
subchannel removal.

\subsection{Comparison of allocations}

To compare the two optimization policies, we call $\mathcal{B}$ the
allocation that maximizes the system margin and $\mathcal{C}$ the
allocation that minimizes the BER.  Table~\ref{tab_robcmp} gives an
example of bit-rate allocation over 20 subchannels where the SNR
follows a Rayleigh distribution and with $\beta=1$.  In this example
the PSDNR is equal to 25~dB and the maximum allowed bit-rate per
subchannel is never reached. As expected, the system margin
minimization leads to a minimal SNR-gap, $\min_i\gamma_i$, higher than
that provided by the BER minimization policy with a gain of 0.3~dB.
On the other hand, the BER minimization policy leads to BER lower than
that provided by system margin minimization ($2.6\cdot10^{-5}$ versus
$3.1\cdot10^{-5}$).  In this example, the dissimilarity is
$\mu(\mathcal{B,C})=0.1$ and two subchannels convey different
bit-rates.  All these results are obtained with $r_\text{max}=10$.

\begin{table}
  \centering
  \caption{Example of system margin and BER with $n=20$, $R=100$, $\mathsf{psdnr}=25~\mathrm{dB}$ and $\beta=1$.}
  \label{tab_robcmp}
  \begin{tabular}{@{}r|cc@{}}
    &system margin&BER\\
    &maximization&minimization\\
    &($\mathcal{B}$)&($\mathcal{C}$)\\
    \hline
    $\min\limits_i\gamma_i$&6.9 dB&6.6 dB\\
    $\mathsf{ber}$&$3.1\cdot10^{-5}$&$2.6\cdot10^{-5}$\\\hline
  \end{tabular}
\end{table}

This example shows that the difference between the two allocation
policies is not so important. The question is whether both
allocations converge and if they converge then in what case. The
following theorem answers the question.

\begin{theorem}\label{theo_cviter}
  In high SNR regime with square QAM and under bit-rate and
  peak-power constraints, the greedy-type allocation that maximizes
  the system margin converges to the greedy-type allocation that
  minimizes the BER.
\end{theorem}

\begin{IEEEproof}
  See Appendix~\ref{ann_cviter}.
\end{IEEEproof}

The consequence of Theorem~\ref{theo_cviter} is that the dissimilarity
between allocation that maximizes the system margin and allocation
that minimizes the BER is null in high SNR regime and with square QAM.
With square QAM, $\beta$ should be a multiple of 2. Note that with
square modulations, $\beta$ can also be equal to 1 if the modulations
are, for example, those defined in ADSL \cite{G9923_02}.
Fig.~\ref{fig_robgapber} exemplifies the convergence with $\beta=2$ as
we will see later in Section~\ref{sec_cmp}.

\section{Optimal analytical allocations}\label{sec_anal}

The analytical method is based on convex optimization theory
\cite{LUO06}.  Unconstrained modulations lead to bit-rates $r_i$
defined in $\mathbb{R}$. With $r_i\in\mathbb{R}_+$ the solution is the
waterfilling one.  With bounded modulation order, i.e.\ $0\leq
r_i\leq r_\text{max}$, the solution is quite different from the
waterfilling one. The solution is obtained in the framework of
generalized Lagrangian relaxation using Karush-Kuhn-Tucker (KKT)
conditions~\cite{BOYD04}.

As the bit-rates are continuous and not only integers, the constraints
\eqref{eq_const} do not hold anymore and become
\begin{align}\label{eq_constr}
  \sum_{i=1}^nr_i=R\quad\forall i\ 0\leq r_i\leq r_\text{max}\ .
\end{align}
The KKT conditions associated to the general problem \eqref{eq_pbgen}
with \eqref{eq_constr} instead of \eqref{eq_const} write~\cite{BOYD04}
\begin{align}
    -r_i&\leq0\,,\ \forall i\in[1,n]\label{eq_kkta}\\
    r_i-r_\text{max}&\leq0\,,\ \forall i\in[1,n]\label{eq_kktb}\\
    R-\sum\limits_{i=1}^nr_i&=0\label{eq_kktc}\\
    \mu_i&\geq0\,,\ \forall i\in[1,n]\label{eq_kktd}\\
    \nu_i&\geq0\,,\ \forall i\in[1,n]\label{eq_kkte}\\
    \mu_ir_i&=0\,,\ \forall i\in[1,n]\label{eq_kktf}\\
    \nu_i(r_i-r_\text{max})&=0\,,\ \forall i\in[1,n]\label{eq_kktg}\\
    \frac{\partial}{\partial r_i}\phi\Big(\{r_j\}_{j=1}^n\Big)-\lambda-\mu_i+\nu_i&=0\,,\ \forall i\in[1,n]\label{eq_kkth}
\end{align}
The first three conditions \eqref{eq_kkta}--\eqref{eq_kktc} represent
the primal constraints, conditions \eqref{eq_kktd} and \eqref{eq_kkte}
represent the dual constraints, conditions \eqref{eq_kktf} and
\eqref{eq_kktg} represent the complementary slackness and condition
\eqref{eq_kkth} is the cancellation of the gradian of Lagrangian with
respect to $r_i$. When the primal problem is convex, i.e.\
$\phi\big(\{r_i\}_{i=1}^n\big)$ is convex and the constraints are
linear, the KKT conditions are sufficient for the solution to be
primal and dual optimal.  For the system margin maximization problem,
the function $\phi(\cdot)$ is convex over all input bit-rates and SNR
whereas this function is no more convex for the BER minimization
problem.  Appendix~\ref{ann_conv} gives the convex domain of the
function $\phi(\cdot)$ in the case of BER minimization problem.

The properties of the studied function $\phi(\cdot)$ are such that
\begin{align}
  \frac{\partial}{\partial
    r_i}\phi\Big(\{r_j\}_{j=1}^n\Big)=\psi_i(r_i)\ .
\end{align}
The optimal solution that solve \eqref{eq_kkta}--\eqref{eq_kkth} is
then~\cite{BOYD04}
\begin{align}\label{eq_rsol}
  r_i^*(\lambda)=\left\{\begin{array}{l@{\quad\text{if}\quad}l}
      0&\lambda\leq\psi_i(0)\\
      \psi_i^{-1}(\lambda)&\psi_i(0)<\lambda<\psi_i(r_\text{max})\\
      r_\text{max}&\lambda\geq\psi_i(r_\text{max})
    \end{array}\right.
\end{align}
for all $i\in[1,n]$ and with $\lambda$ verifying the constraint
\begin{align}\label{eq_rlambda}
  \sum_{i=1}^nr_i^*(\lambda)=R\ .
\end{align}

It is worthwhile noting that the above general solution is the
waterfilling one if $r_\text{max}\geq R$. The waterfilling is also the
solution in the following case. Let $\mathcal{I}'$ the subset such
that $\forall i\in\mathcal{I}'$, $r_i^*\notin\{0,r_\text{max}\}$ and
let $R'$ the target bit-rate over $\mathcal{I}'$.  In this subset,
$\{r_i^*\}_{i\in\mathcal{I}'}$ are solutions of
\begin{align}\label{eq_gsol}
  \left\{\begin{array}{r@{\ =\ }l}
    \frac{\partial}{\partial r_i}\phi\Big(\{r_j\}_{j=1}^n\Big)-\lambda&0\,,\ \forall i\in\mathcal{I}'\\
    R'-\sum\limits_{i\in\mathcal{I}'}r_i(\lambda)&0
  \end{array}\right.
\end{align}
This is the solution of \eqref{eq_pbgen} with unbounded modulations
over the subchannel index subset $\mathcal{I}'$. If
$\mathcal{I}'=\{1,\cdots,n\}$ and $R'=R$, \eqref{eq_gsol} is also the
solution of \eqref{eq_pbgen} with unconstrained modulations.

\subsection{System margin maximization}

\begin{theorem}\label{theo_gap}
  Under bit-rate and peak-power constraints, the asymptotic bit
  allocation with unconstrained modulations which minimizes the inverse
  system margin is given by
  \[\forall i\in[1,n],\quad
  r_i^*=\frac{R}{n}+\frac{1}{n}\sum_{j=1}^n\log_2\frac{\mathsf{snr}_i}{\mathsf{snr}_j}\]
\end{theorem}
\begin{IEEEproof}
  See Appendix~\ref{ann_gap}.
\end{IEEEproof}

The solution given by Theorem~\ref{theo_gap} holds for high modulation
orders which defines the asymptotic regime, cf.
Appendix~\ref{ann_gap}. As the modulations are unconstrained, the
bit-rates $\{r_i^*\}_{i=1}^n$ are real numbers defined in
$\mathbb{R}$. With bounded modulation orders, Theorem \ref{theo_gap}
is used into \eqref{eq_rsol} to solve \eqref{eq_pbgen} with
constraints \eqref{eq_constr}. If the set $\mathcal{I}'$ is known,
then Theorem~\ref{theo_gap} can be used directly to allocate the
subchannel bit-rates. Otherwise $\mathcal{I}'$ should be found first.

The expression of $r_i^*$ in Theorem~\ref{theo_gap} is a function of
the target bit-rate $R$, the number $n$ of subchannels and the ratios
of SNR. This expression is independent of mean received SNR or PSDNR.
It does not depend on link budget but only on relative distribution of
subchannel coefficients $\{|h_i|^2\}_{i=1}^n$.

\subsection{BER minimization}

The arithmetic mean BER minimization has been analytically solved for
example in \cite{PALO03, ISER04}. This arithmetic mean measure needs
to employ the same number of bits per constellation which limits the
system efficiency. The following theorem gives the solution of the
weighted mean BER minimization that allows variable constellation
sizes in the multichannel system.

\begin{theorem}\label{theo_lag}
  Under bit-rate and peak-power constraints, the asymptotic bit
  allocation with unconstrained QAM which minimizes the BER is given
  by
  \[\forall i\in[1,n],\quad r_i^*=\frac{R}{n}+\frac{1}{n}\sum_{j=1}^n\log_2
  \frac{\mathsf{snr}_i}{\mathsf{snr}_j}\] with equal in-phase and
  quadrature bit-rates.
\end{theorem}
\begin{IEEEproof}
  See Appendix~\ref{ann_lag}.
\end{IEEEproof}

The solution given by Theorem~\ref{theo_lag} holds for high modulation
orders and for subchannel BER lower than 0.1, and these parameters
define the asymptotic regime in this case, cf. Appendix~\ref{ann_lag}.
The optimal asymptotic allocation leads to square QAM with
$\sqrt{r_i^*}$ conveyed bit-rate in each in-phase and quadrature
components of the signal of subchannel $i$. With bounded modulation
orders, Theorem~\ref{theo_lag} is used in \eqref{eq_rsol} to solve
\eqref{eq_pbgen} with constraints \eqref{eq_constr}. It is important
to note that in asymptotic regime, BER minimization and system margin
maximization lead to the same subchannel bit-rate allocation. In that
case, the asymptotic regime is defined by the more stringent context
which is the BER minimization.  As we will see in
Section~\ref{sec_cmp}, this asymptotic behavior can be observed when
$\beta=2$.

The main drawback of the formulas in Theorem~\ref{theo_lag} and
Theorem~\ref{theo_gap} is that the bit-rates $\{r_i^*\}_{i=1}^n$ are
expressed in $\mathbb{R}$ and not in $\mathbb{R}_+$. To find the set
$\mathcal{I}'$, the negative subchannel bit-rates and those higher
than $r_\text{max}$ should be clipped and $\mathcal{I}'$ can be found
iteratively \cite{BACC02}. But clipping negative bit-rates first can
decrease those higher than $r_\text{max}$ and clipping bit-rates
higher than $r_\text{max}$ first can increase the negative ones. It is
then not possible to apply first the waterfilling solution and after
that to clip the bit-rates $r_i$ greater than $r_\text{max}$ to
converge to the optimal solution. Finding the set $\mathcal{I}'$
requires many comparisons and we propose a fast iterative solution
based on generalized secant method.

\subsection{Lagrangian resolution}

To solve \eqref{eq_rsol}, numerical iterative methods are required. It
is important to observe that the function defined in \eqref{eq_rsol}
is not differentiable and thus, methods like Newton's cannot be
used~\cite{BACC02}.  We use the proposed generalized secant method to
better fit the function-depending weight and increase the speed of the
convergence.  An important point for the iterative method is that the
initialization must embrace all possible scenarios and should be as
close as possible to the final solution.

\begin{figure}
  \centering
  \includegraphics[width=\linewidth]{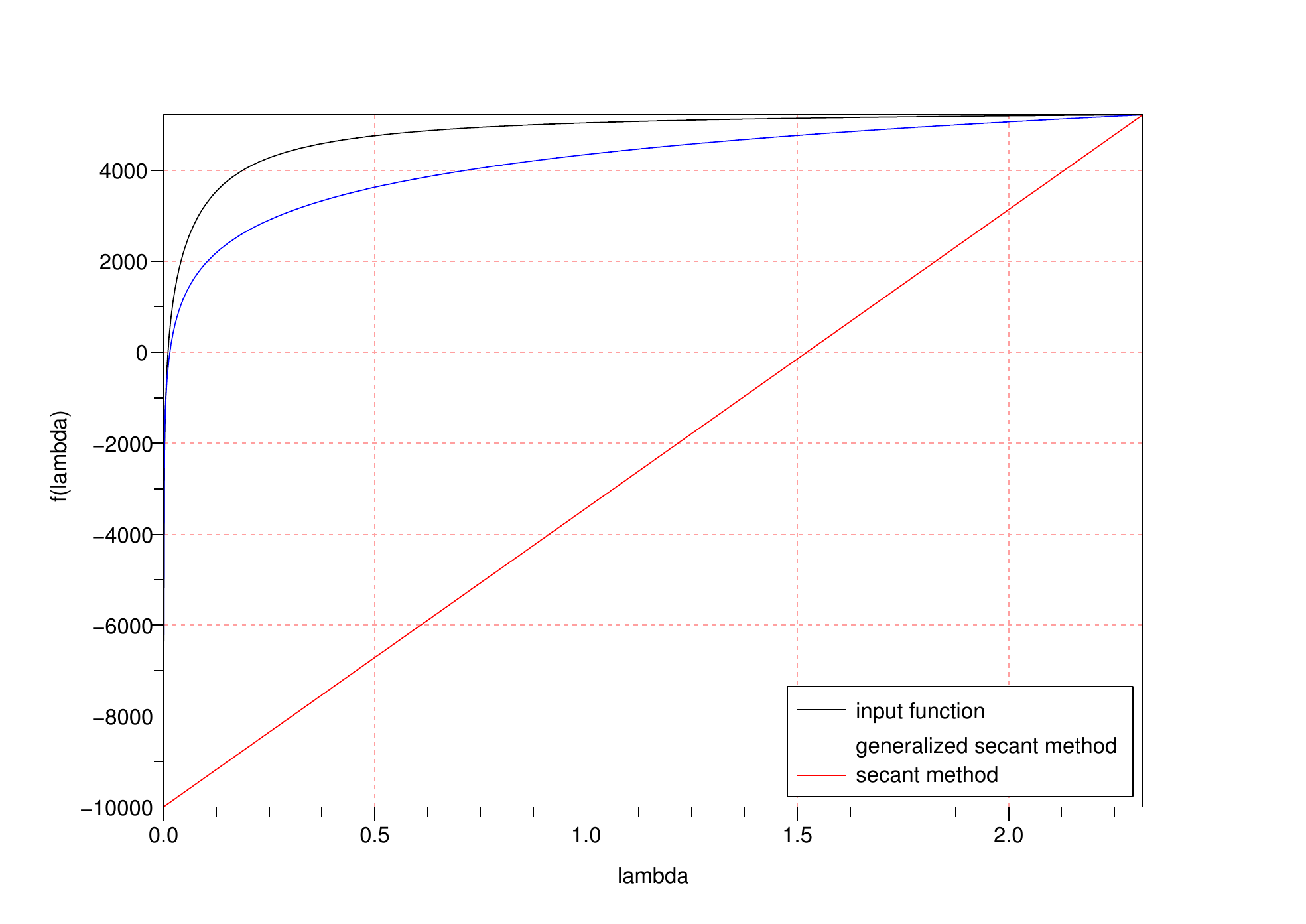}
  \caption{Approximation of input function $f(\lambda)$ with
    generalized secant method and secant method, $n=1024$ and
    $r_\text{max}=15$.}
  \label{fig_shape}
\end{figure}

The root of the function defined by \eqref{eq_rlambda} is now
calculated.  Let
\begin{align}
  f(\lambda)=\sum_{i=1}^nr_i(\lambda)-R\,.
\end{align}
Theorems \ref{theo_gap} and \ref{theo_lag} show that $r(\lambda)$ is
the sum of $\log_2(\cdot)$ functions. This is the reason why the
function $\log_2(\cdot)$ is used in generalized secant method.
Fig.~\ref{fig_shape} shows three functions versus the parameter
$\lambda$.  The first function is the input function $f(\lambda)$, the
second one is the function used by the generalized secant method, and
the last one if the linear function used by the secant method.  In
this example, the common points are $\lambda=0$ and $\lambda=2.3$.  As
it is shown, the generalized secant method better fit the input
function than the secant method and therefore can improve the speed of
the convergence to find the root which is around $\lambda=1/80$ in
this example.

To ensure the non empty root of the input function $f(\lambda)$, the
secant methods should be initialized with $\lambda_1$ and $\lambda_2$
such as $f(\lambda_1)<0$ and $f(\lambda_2)>0$. For both optimization
problems, system margin maximization and BER minimization, the
parameter $\lambda$ is given by the function $\psi_i(r_i)$ and it can
be reduced to $\lambda=\frac{2^{r_i}}{\mathsf{snr}_i}$, as shown in
appendices \ref{ann_gap} and \ref{ann_lag}. Parameters
$\{\lambda_1,\lambda_2\}$ are then chosen as
\begin{align}
  \lambda_1=\frac{1}{\max\limits_i\mathsf{snr_i}}\quad\text{and}\quad
  \lambda_2=\frac{2^{r_\text{max}}}{\min\limits_i\mathsf{snr_i}}\,.
\end{align}
Using \eqref{eq_rsol}, $\lambda\leq\lambda_1$ leads to
$r_i(\lambda)=0$ for all $i$, and $\lambda\geq\lambda_2$ leads to
$r_i(\lambda)=r_\text{max}$ for all $i$. Then, it follows that
$f(\lambda_1)<0$ and $f(\lambda_2)>0$ if $R\in(0,nr_\text{max})$.

\begin{figure}
  \centering
  \includegraphics[width=\linewidth]{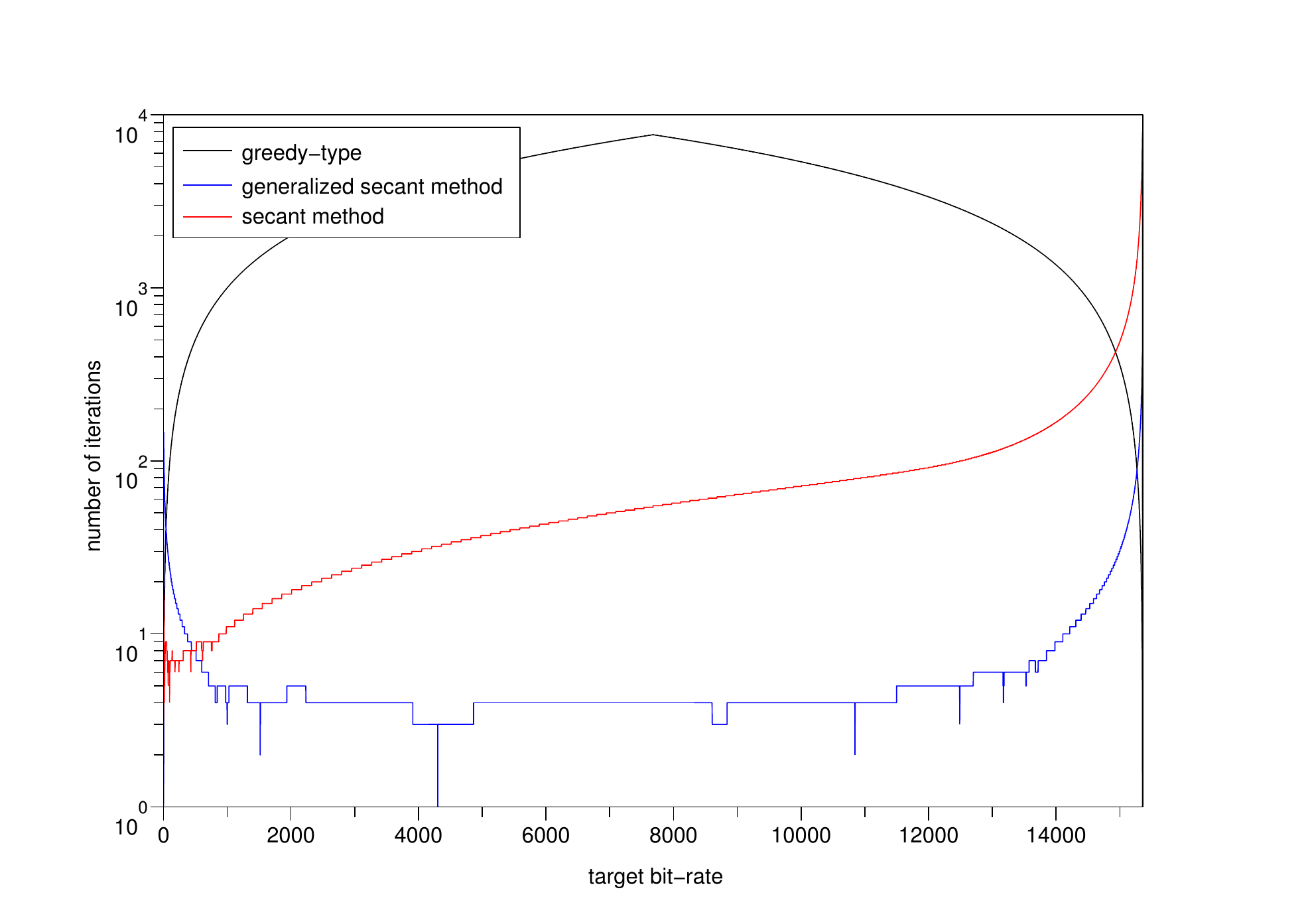}
  \caption{Number of iterations of secant and generalized secant
    methods, and greedy-type algorithm versus the target bit-rate,
    $n=1024$, $r_\text{max}=15$.}
  \label{fig_nbiterlag}
\end{figure}

Fig.~\ref{fig_nbiterlag} shows the needed number of iterations for the
convergence of the generalized and conventional secant methods versus
the target bit-rate $R$. Results are given over a Rayleigh
distribution of the subchannel SNR with 1024 subchannels.  The
possible bit-rates are then $R\in[0,n\times r_\text{max}]$ and
$\beta=2$. Here, $r_\text{max}=15$ and then $R\leq15360$ bits per
multidimensional symbol. For comparison, the number of iterations
needed by the greedy-type algorithm is also plotted.  Note that the
greedy-type algorithm can start by empty bit-rate or by full bit-rate
limited by $r_\text{max}$ for each subchannel. The number of
iterations is then given by $\min\{R,nr_\text{max}-R\}$. The iterative
secant and generalized secant methods are stopped when the bit-rate
error is lower than 1. A better precision is not necessary since exact
bit-rates $\{r_i\}_{i\in\mathcal{I}'}$ can be computed using Theorems
\ref{theo_gap} and \ref{theo_lag} when $\mathcal{I}'$ is known. As it
is shown in Fig.~\ref{fig_nbiterlag}, the generalized secant method
converges faster than the secant method, except for the very low
target bit-rates $R$. For very high target bit-rates, near from
$n\times r_\text{max}$, generalized secant method can lead to a number
of iterations higher than those of greedy-type algorithm. Except for
these particular cases, the generalized secant method needs no more
than 4--5 iterations to converge. In conclusion, we can say that with
Rayleigh distribution of $\{\mathsf{snr}_i\}_{i=1}^n$, the generalized
secant method converges faster than secant method or greedy-type
algorithm for target bit-rate $R$ such that
$3\%\leq\frac{R}{nr_\text{max}}\leq97\%$.

Using generalized secant method the bit-rates are not integer and for
all $i$, $r^*_i\in[0,r_\text{max}]$. These solutions have to be
completed to obtain integer bit-rates.

\subsection{Integer-bit solution}

Starting from continuous bit-rates allocation previously presented, a
loading procedure is developed taking into account the integer nature
of the bit-rates to be allotted. A simple solution is to consider the
integer part of $\{r_i^*\}_{i\in\mathcal{I}'}$ and to complete by a
greedy-type algorithm to achieve the target bit-rate $R$. The integer
part of $\{r_i^*\}_{i\in\mathcal{I}'}$ is then used as a starting
point for the greedy algorithm. This procedure can lead to a high
number of iterations, therefore the secant or bisection methods are
suitable to reduce the complexity.  The problem to solve is then to
find the root of the following function \cite{BACC02}
\begin{align}\label{eq_hrn}
  g(\alpha)=\sum_{i\in\mathcal{I}'}\lfloor r_i^*+\beta\alpha\rfloor-R'\,,
\end{align}
where $r_i^*$, $\mathcal{I}'$ and $R'$ are given by the continuous
Lagrangian solution. This is a suboptimal integer bit-rate problem and
the optimal one needs to find $\{\alpha_i\}_{i=1}^n$ instead of a
unique $\alpha$. As the optimal solution leads to a huge complexity,
it is not considered.  The function \eqref{eq_hrn} is a non decreasing
and non differentiable staircase function such that $g(0)<0$, $g(1)>0$
because $\sum_{i\in\mathcal{I}'}r_i^*=R'$. The iterative methods can
then be initialized with $\alpha_1=0$ and $\alpha_2=1$.

\begin{figure}
  \centering
  \includegraphics[width=\linewidth]{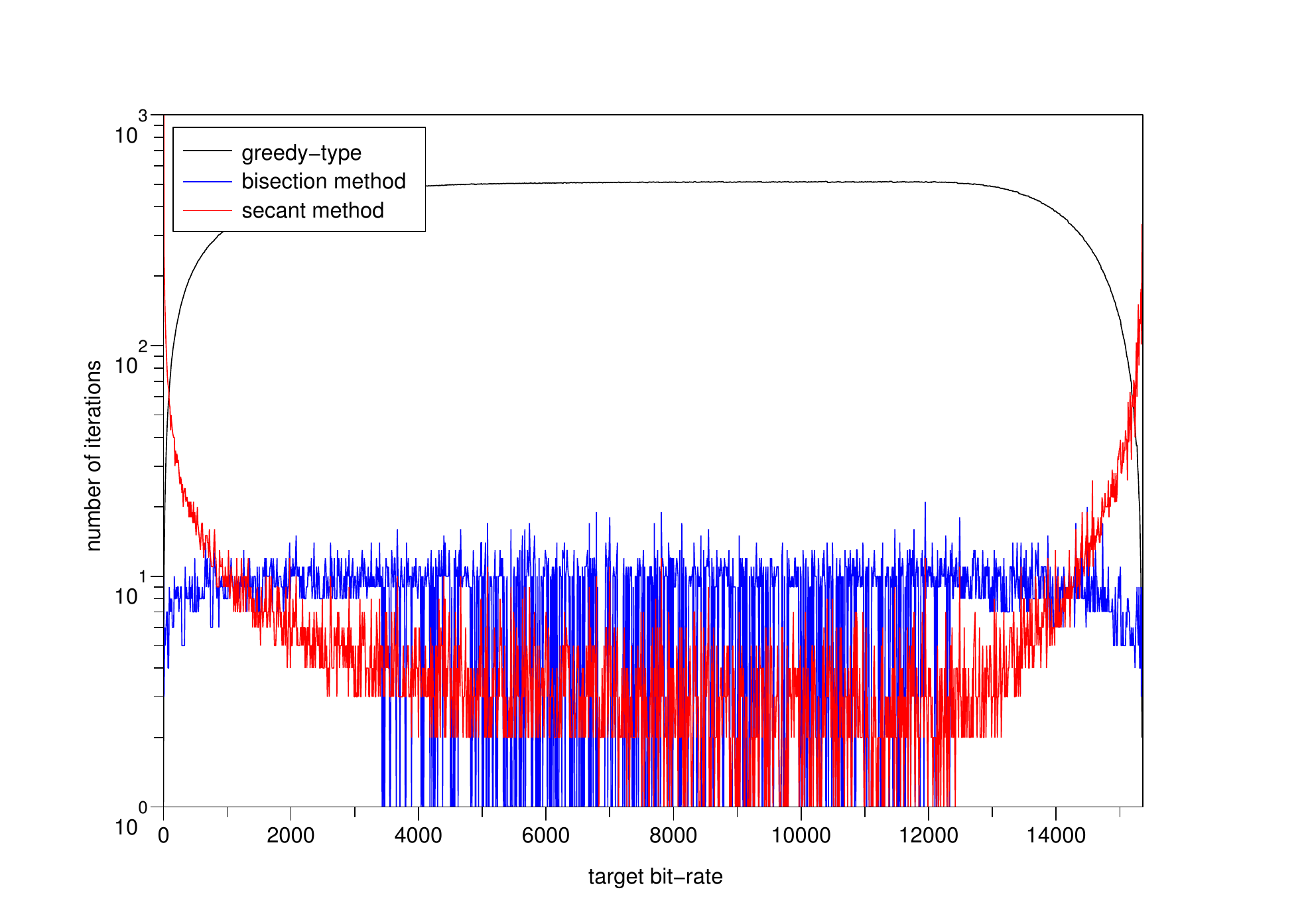}
  \caption{Number of iterations of bisection and secant methods, and
    greedy-type algorithm for integer-bit solution versus target
    bit-rate, $n=1024$, $r_\text{max}=15$.}
  \label{fig_nbiteral}
\end{figure}

Two iterative methods are compared, the bisection one and the secant
one. Both methods are also compared to the greedy-type algorithm.
Fig.~\ref{fig_nbiteral} presents the number of iterations of the three
methods to solve the integer-bit problem of the Lagrangian solution
with $\beta=1$.  Results are given over a Rayleigh distribution of the
subchannel SNR, with 1024 subchannels and the target bit-rates are
between 0 and $n\times r_\text{max}=15360$.  As it is shown, the
convergence is faster with bisection method than with greedy-type
algorithm. For target bit-rates between 10\% and 90\% of the maximal
loadable bit-rate, the secant method outperforms the bisection one
with a mean number of iterations around 4 whereas the number of
iterations for bisection method is higher than 8.
Fig.~\ref{fig_nbiteral} also shows that $|g(0)|$ is all the time lower
than the half of number of subchannels and around this value for
target bit-rate between 10\% and 90\% of the maximal loadable
bit-rate. Then, if the complexity induces by the greedy-type algorithm
to solve the integer-bit problem of the Lagrangian solution is
acceptable in a practical communication system, this greedy-type
completion can be used and leads to the optimal allocation. This
result obtained without proof means that the greedy-type procedure has
enough bits to converge to the optimal solution. If the number of
iterations induced by the greedy-type algorithm is too high (this
number is around $n/2$), the secant method can be used.

The overall analytical resolution of \eqref{eq_pbgen} needs few
iterations and its complexity is $\mathcal{O}(n)$ instead of
$\mathcal{O}(Rn)$ for the optimal greedy-type algorithm. Whereas the
continuous solution of \eqref{eq_pbgen} is optimal, the analytical
integer bit-rate solution is suboptimal.

\section{Greedy-type versus analytical allocations}\label{sec_cmp}

In the previous section, algorithm complexities have been compared.
In this section, robustness comparison is presented and the analytical
solutions obtained in asymptotic regime are also applied in non
asymptotic regime which means that $\beta=1$ and modulation orders can
be low.

\begin{figure*}
  \centering
  \includegraphics[width=\linewidth]{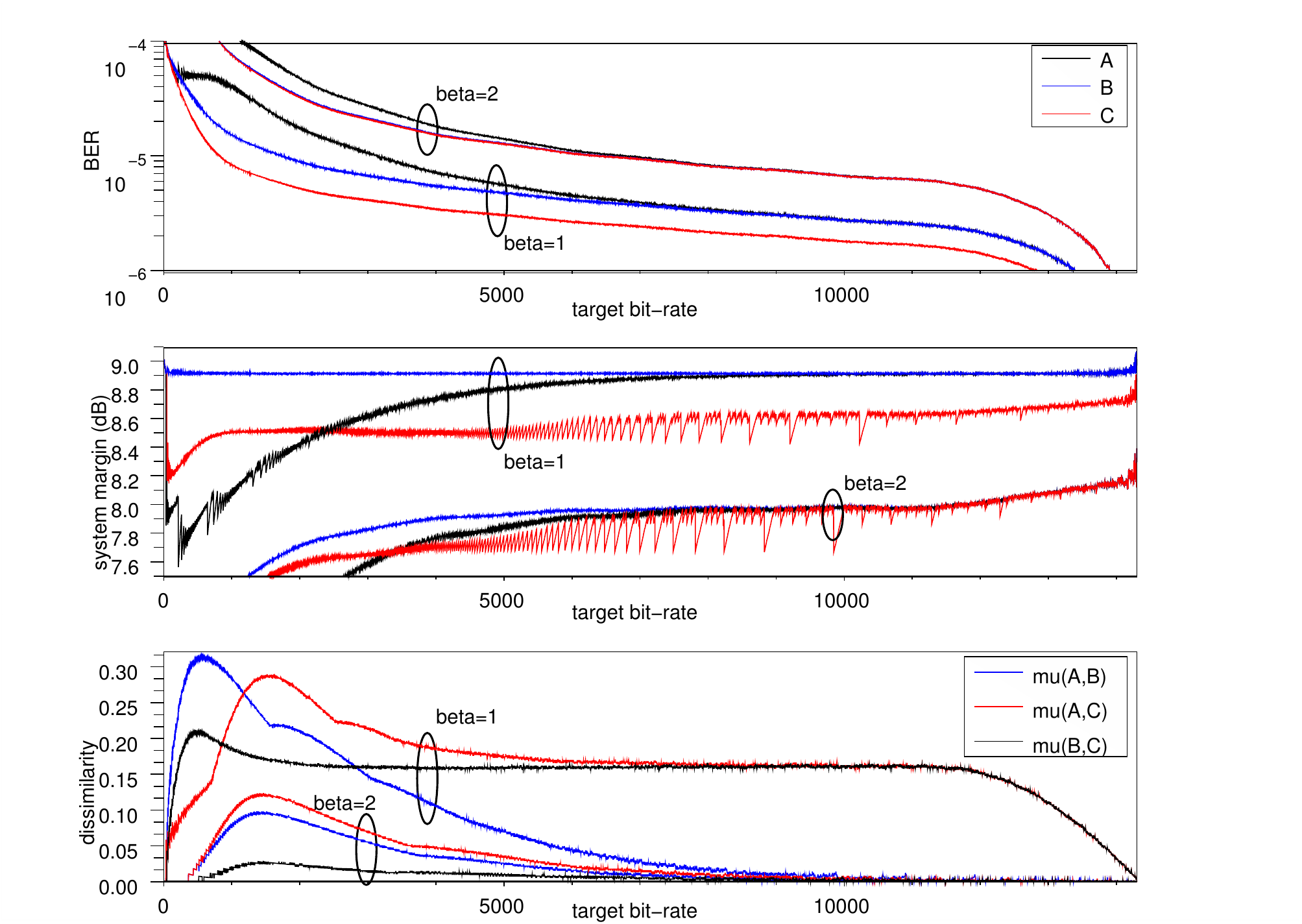}
  \caption{BER, system margin and dissimilarity versus target bit-rate
    for Lagrangian ($\mathcal{A}$), greedy-type system margin
    maximization ($\mathcal{B}$) and greedy-type BER minimization
    ($\mathcal{C}$) algorithms, $n=1024$, $r_\text{max}=14$,
    $\beta\in\{1,2\}$.}
  \label{fig_robgapber}
\end{figure*}

Fig.~\ref{fig_robgapber} presents the output BER and the system margin
of three allocation policies versus the target bit-rate $R$. The first
one, $\mathcal{A}$, is obtained using analytical optimization, the
second, $\mathcal{B}$, is the solution of the greedy-type algorithm
which maximizes the system margin and the third, $\mathcal{C}$, is the
solution of the greedy-type algorithm which minimizes the BER. Two
cases are presented, one with $\beta=1$ and the other with $\beta=2$.
All subchannel BER are lower than $2\cdot10^{-2}$ to use valid BER
approximations. To compare the case $\beta=1$ with the case $\beta=2$,
$r_\text{max}$ is equal to 14.  Results are given over a Rayleigh
distribution of the subchannel SNR, with 1024 subchannels. Note that
with $\beta=1$, the system margin of allocation $\mathcal{B}$ is
almost equal to 8.9~dB for all target-bit rates. This constant system
margin $\gamma$ is not a feature of the algorithm but is only a
consequence of the relation between the target bit-rate and the PSDNR.

To enhance the equivalences and the differences between the allocation
policies, the dissimilarity is also given in Fig.~\ref{fig_robgapber}
with $\beta=1$ and $\beta=2$. As expected and in both cases $\beta=1$
and $\beta=2$, the minimal BER are obtained with allocation
$\mathcal{C}$, and the maximal system margins with allocation
$\mathcal{B}$. With $\beta=1$ and when the target bit-rate increases,
the Lagrangian solution converges faster to the optimal system margin
maximization solution $\mathcal{B}$ than to the optimal BER
minimization solution $\mathcal{C}$. Note that Theorem~\ref{theo_lag}
is an asymptotic result valid for square QAM. With $\beta=1$, the QAM
can be rectangular and the asymptotic result of Theorem~\ref{theo_lag}
is not applicable, contrary to the result of Theorem~\ref{theo_gap}
where there is not any condition on the modulation order.  With
$\beta=2$, the modulation granularity is lower and then the output BER
are higher than those obtained with $\beta=1$, as the output system
margins are lower than those obtained with $\beta=1$.  This case
$\beta=2$ shows the equivalence between the optimal system margin
maximization allocation and the optimal BER minimization allocation.
In this case, the asymptotic result given by Theorem~\ref{theo_cviter}
and \ref{theo_lag} can be applied because the modulations are square
QAM, and the convergence is ensured with high modulation orders, i.e.
high target bit-rates. Beyond a mean bit-rate per subchannel around
10, that corresponds to a target bit-rate around $10^4$, all the
allocations $\mathcal{A}$, $\mathcal{B}$ and $\mathcal{C}$ are
equivalent and the dissimilarity is almost equal to zero. In non
asymptotic regime, the differences in BER and system margin are low.
The system margin differences are lower than 1~dB, and the ratios
between two BER are around 3. In practical integrated systems these
low differences will not be significant and will lead to similar
solutions for both optimization policies. Therefore, these allocation
can be interchanged.

\section{Application: DMT for power line channel}\label{sec_sim}

\begin{figure}
  \centering
  \includegraphics[width=\linewidth]{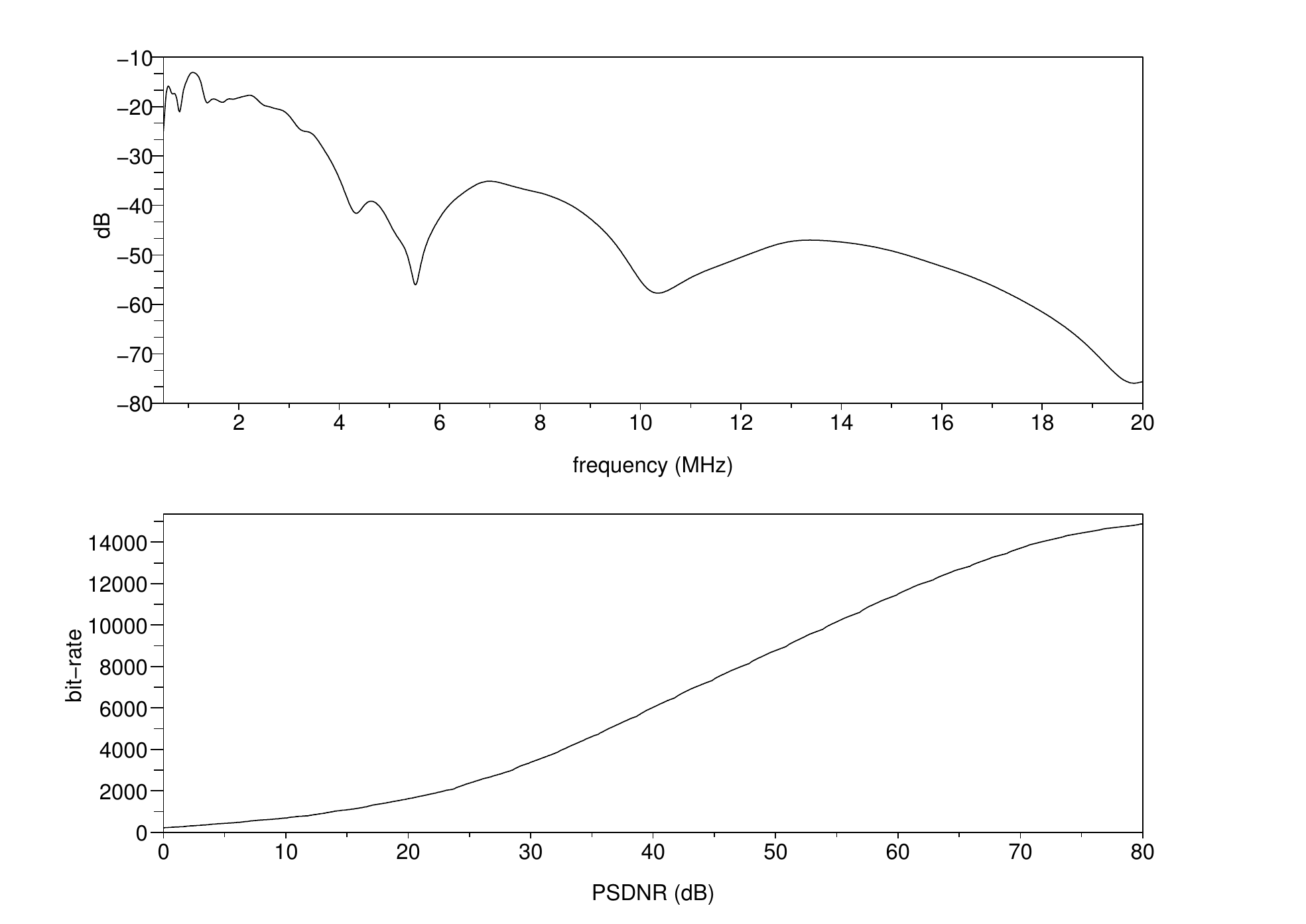}
  \caption{Frequency response of 15 paths PLC model~\cite{ZIMM00}
    versus frequency and target bit-rate versus input PSDNR.}
  \label{fig_zimm}
\end{figure}

All previous results are presented over Rayleigh distribution of
subchannel SNR. This distribution can occur in PLC channel
\cite{TLIC08b} but distribution with higher SNR range is also possible
\cite{ZIMM00}. The robustness is evaluated in harsh condition and the
model proposed in \cite{ZIMM00} is then chosen. Fig.~\ref{fig_zimm}
shows the frequency response of 15 paths PLC model in $[0.5;20]$~MHz
bandwidth where the subchannels undergo more than 60~dB SNR range. In
PLC systems, the subchannels are the subcarriers of the multicarrier
system and the PLC communication system uses 1024 subcarriers,
$r_\text{max}=15$ as in xDSL communication systems and $\beta=1$. All
QAM constellations between 4-QAM to 32768-QAM plus BPSK are used. Note
that with 60~dB of SNR range, there are possible $r_i=0$ and $r_j=15$
in the same multicarrier symbol since around 45~dB of SNR range is
sufficient to have both $r_i=0$ and $r_j=15$ in the same multicarrier
symbol. The robustness measures are evaluated for different target
bit-rates which are given with the following arbitrary equation
\begin{align}
  R=\left\lfloor\sum_{i=1}^n\min\Big(\log_2\big(1+\frac{\mathsf{snr}_i}{2}\big),r_\text{max}\Big)\right\rfloor\,.
\end{align}
This equation ensures reliable communications for all the input
target bit-rates or PSDNR. The empirical relationship between PSDNR
and target bit-rate is also given in Fig.~\ref{fig_zimm}.

\begin{figure*}
  \centering
  \includegraphics[width=.85\linewidth]{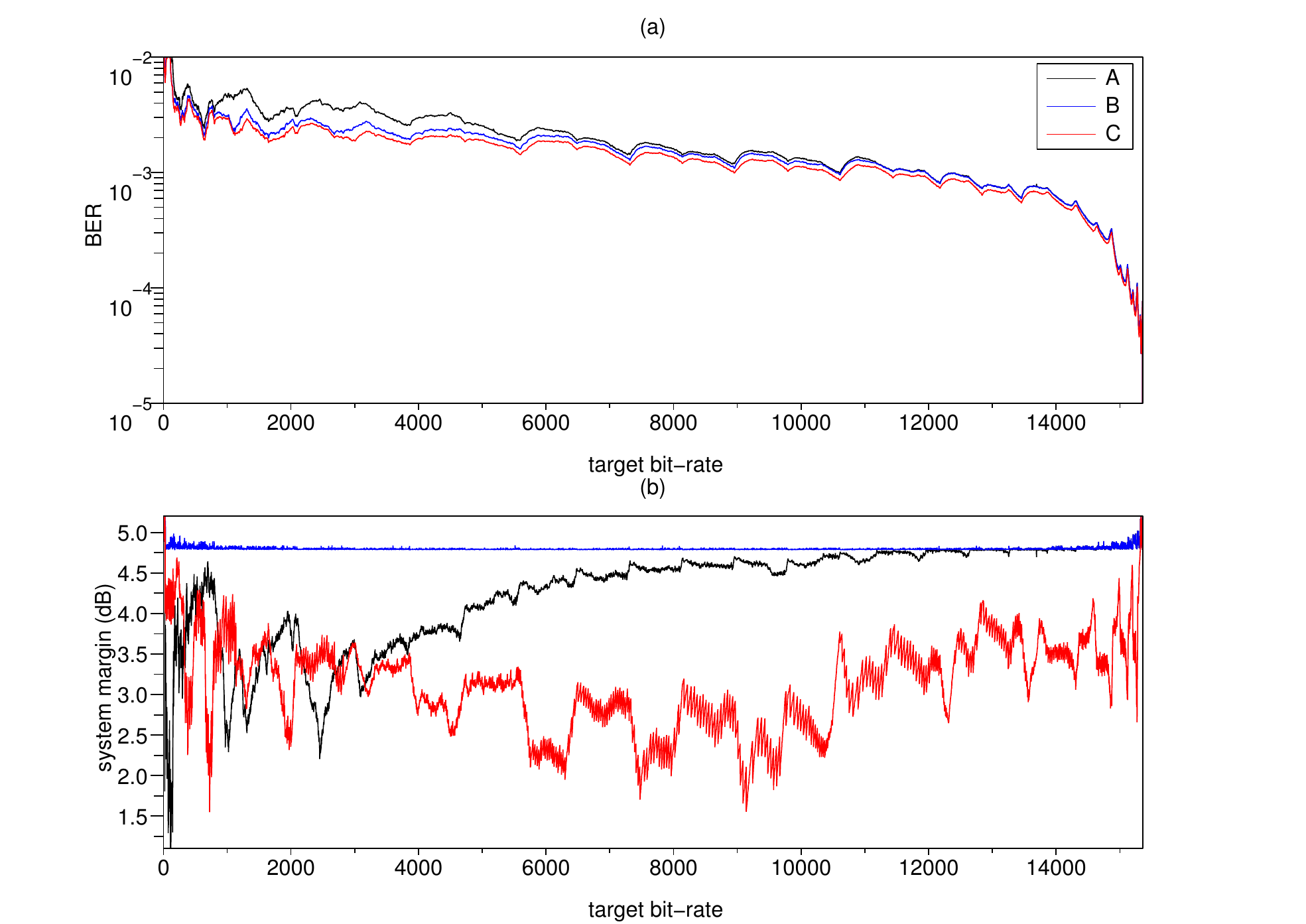}
  \includegraphics[width=.85\linewidth]{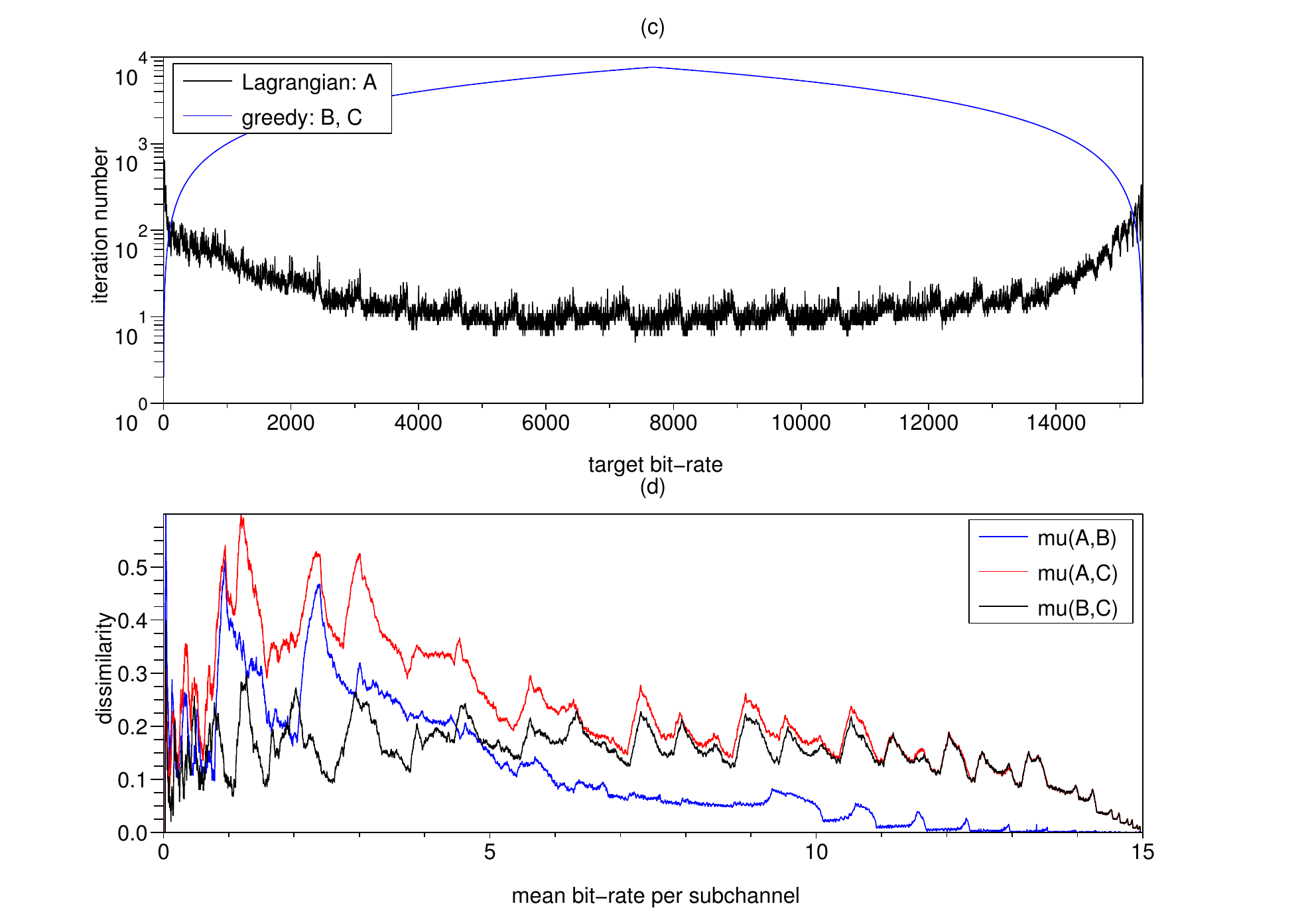}
  \caption{(a) BER, (b) system margin versus target bit-rate for
    Lagrangian ($\mathcal{A}$), greedy-type system margin maximization
    ($\mathcal{B}$) and greedy-type BER minimization ($\mathcal{C}$)
    algorithms, (c) iteration number of Lagrangian solution and
    greedy-type algorithms versus target bit-rate, and (d)
    dissimilarities $\mu(\mathcal{A,B})$, $\mu(\mathcal{A,C})$ and
    $\mu(\mathcal{B,C})$ versus mean bit-rate per subchannel,
    $n=1024$, $r_\text{max}=15$.}
  \label{fig_zimmb}
\end{figure*}

Fig.~\ref{fig_zimmb} sums up the characteristics and the results of
three allocation policies in PLC context.  The first allocation,
$\mathcal{A}$, is obtained using analytical optimization developed in
Section~\ref{sec_anal}, the second, $\mathcal{B}$, is the solution of
the greedy-type algorithm given by Lemma~\ref{theo_itergap} which
maximizes the system margin and the third, $\mathcal{C}$, is the
solution of the greedy-type algorithm given by Lemma~\ref{theo_iter}
which minimizes the BER.  In this figure, the BER, the system margin
in decibel and the needed number of iterations are plotted versus the
input target bit-rates, subfigure (a), (b) and (c) respectively.  The
subfigure (d) gives the dissimilarity measure versus the mean bit-rate
per subchannel which is equal to $R/n$. In subfigure (a) the minimal
BER is obtained with allocation $\mathcal{C}$, as expected. The
allocation $\mathcal{A}$ offers the worse BER with a maximal ratio of
2 compared to the best result. For high bit-rates this ratio converges
to~1. In subfigure (b) the maximal system margin is given by
allocation $\mathcal{B}$, as expected,.  The maximal difference
between allocations $\mathcal{B}$ and $\mathcal{C}$ is around 3~dB.
The number of iterations needed with Lagrangian method, allocation
$\mathcal{A}$, is very low compared with optimal greedy-type ones,
allocations $\mathcal{B}$ and $\mathcal{C}$.  Subfigure (c) shows that
only 20 iterations are sufficient for most cases to perform allocation
with method $\mathcal{A}$, when up to 7680 iterations are needed with
greedy-type algorithms. This maximal number of iterations is given by
$\frac{n}{2}r_\text{max}$. The dissimilarity, subfigure (d), is very
high for low mean bit-rate per subchannel. When the mean number of
bits per constellation is low, the asymptotic results do not hold.
This dissimilarity decreases with the increase of the mean bit-rate
per subchannel where the asymptotic results become valid. Note that as
the PSDNR varies with the target bit-rate, the BER curves are not
simply decreasing and the system margin curves are not simply
increasing with the increase of the target bit-rate.

All these results show that the allocation $\mathcal{A}$ based on
Lagrangian method converges faster to allocation $\mathcal{B}$ (than
to allocation $\mathcal{C}$) when the bit-rate increases. In
communication systems with $\beta=1$, the asymptotic conditions of
analytical solution of minimal BER problem do not hold. Nevertheless,
the difference between allocations $\mathcal{A}$ and $\mathcal{C}$
remains low. The system margin differences are up to 3.5~dB, and the
BER ratios are up to 2.2. The allocation policy should then be
selected depending on the considered robustness measure.

Fig.~\ref{fig_robgapber} and \ref{fig_zimmb}, and complementary
simulations not presented here, show that the system margin difference
between two allocation policies increases when the average value of
the system margin decreases, whereas the BER difference, or ratio,
between two allocation policies decreases when the average value of
system BER increases. With optimized uncoded systems, the case
presented in Fig.~\ref{fig_zimmb} is generally considered for operating
points which correspond to higher BER. Both allocation policies then
lead to similar BER performance but lead to significantly different
system margin performance.

\begin{figure}
  \centering
  \includegraphics[width=\linewidth]{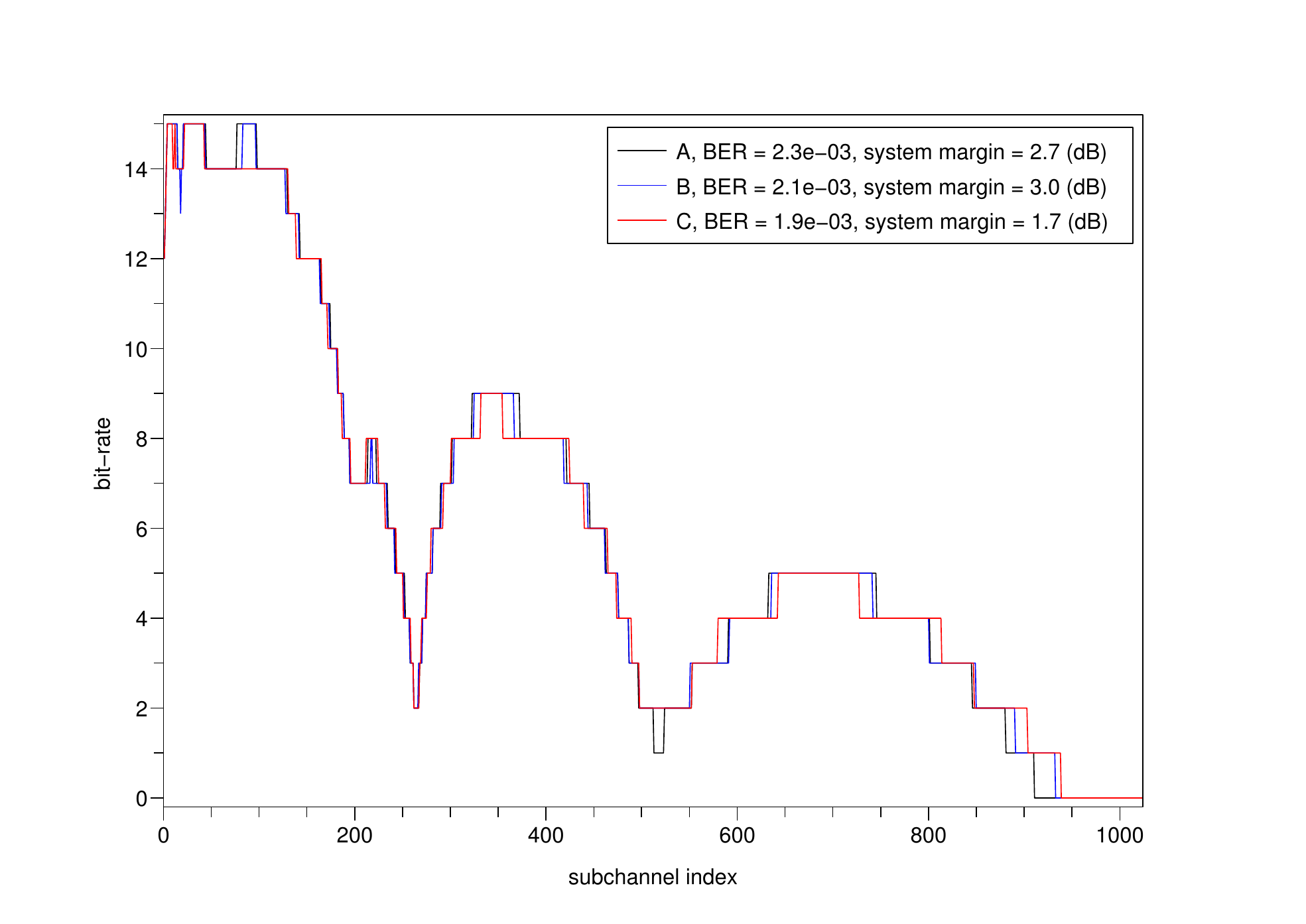}
  \caption{Example of bit-rate allocation versus subchannel index for
    Lagrangian ($\mathcal{A}$), greedy-type system margin maximization
    ($\mathcal{B}$) and greedy-type BER minimization ($\mathcal{C}$)
    algorithms, $n=1024$, $r_\text{max}=15$, $\mathsf{psdnr}=40$~dB,
    $R=6030$.}
  \label{fig_zimmc}
\end{figure}

Fig.~\ref{fig_zimmc} gives an example of bit-rate distribution between
subchannels with an input bit-rate of 6030 bits per multicarrier
symbol which corresponds to a PSDNR of 40~dB. All the bit-rate
distributions follow the shape of the PLC channel frequency response
and exhibit some variations. In this example, the system margin of the
allocation $\mathcal{A}$ is close to the optimal one, and is between
the system margin of the two greedy-type solutions. Allocation
$\mathcal{A}$ offers the highest BER but only 1.2 times higher than
the optimal BER. The main difference between allocations $\mathcal{B}$
and $\mathcal{C}$ is the distribution of even and odd number of bits
per constellation. With allocation $\mathcal{B}$ there are almost as
many odd elements as there are even, whereas allocation $\mathcal{C}$
favors even elements. This behavior of allocation $\mathcal{C}$ is
exemplified in Fig.~\ref{fig_zimmc} with $r_i=4$ and subchannel index
$i$ around 600 and 800, or with $r_i=8$ and $i$ around 400. The
explanation is given by the BER curves in Fig.~\ref{fig_ber} where the
distance between two curves depends on the parity of the number of
bits per constellation. The distance between the curves corresponding
to $r_i=3$ and $r_i=4$ is lower than the distance for the case $r_i=4$
and $r_i=5$. Then, for the same slope of the input frequency response
function, the staircase function of the allocation $\mathcal{C}$ in
Fig.~\ref{fig_zimmc} has larger step size when the numbers of bits per
constellation are even.

All these results in PLC context show that both allocation policies,
system margin maximization and BER minimization, lead to different
bit-rate allocations. These unique allocations have significant system
margin difference whereas the BER difference between these allocations
is very low.  The suboptimal analytical solution based on Lagrangian
offers a good trade-off between performance and complexity.  With
channel coding, not taken into account in this paper, the analysis
remains valid: the ultimate output coded BER is strongly related to
the uncoded BER, and the system margin optimization is independent of
channel coding type.

\section{Conclusion}\label{sec_conc}

Two robustness optimization problems have been analyzed in this paper.
Weighted mean BER minimization and minimal subchannel margin
maximization have been solved under peak-power and bit-rate
constraints.  The asymptotic convergence of both robustness
optimizations has been proved for analytical and algorithmic
approaches. In non asymptotic regime the allocation policies can be
interchanged depending on the robustness measure and the operating
point of the communication system.  We have also proved that the
equivalence between SNR-gap maximization and power minimization in
conventional MM problem does not hold with peak-power limitation
without additional conditions.  Integer bit solution of analytical
continuous bit-rates has been obtained with a new generalized secant
method, and bit-loading solutions have been compared with a new
defined dissimilarity measure.  The low computational effort of the
suboptimal solution, based on the analytical approach, leads to a good
trade-off between performance and complexity.

%%% Local Variables: 
%%% mode: latex
%%% TeX-master: "article"
%%% End: 

%% file: appendix.tex
\appendices

\section{Proof of Lemma \ref{theo_itergap}}\label{ann_itergap}

We prove that the optimal allocation is reached starting from empty
loading with the same intermediate loading than starting from optimal
loading to empty loading. To simplify the notation and without loss of
generality, $\beta=1$.

Let $[r_1^*,\cdots,r_n^*]$ be the optimal allocation that minimizes
the inverse system margin $\gamma(R^*)^{-1}$ for the target bit-rate
$R^*$, then
\begin{align}
  \gamma(R^*)^{-1}=\max_i\frac{2^{r_i}-1}{\mathsf{snr}_i}\ .
\end{align}
Let $[r_1,\cdots,r_n]$ be the optimal allocation that minimizes the
inverse system margin $\gamma(R+1)^{-1}$ for the target bit-rate
$R+1\leq R^*$.  The optimal allocation for target bit-rate $R$ is
obtained iteratively by removing one bit at a time from the subchannel
$k$ with the highest inverse system margin \cite{ZHU07}
\begin{align}
  k=\arg\max_i\frac{2^{r_i}-1}{\mathsf{snr}_i}
\end{align}
or
\begin{align}\label{eq_kup1ann}
  \forall i\in[1,n]\quad\frac{2^{r_k}-1}{\mathsf{snr}_k}\geq
  \frac{2^{r_i}-1}{\mathsf{snr}_i}\ .
\end{align}
The last bit removed is from the subchannel with the lowest
inverse-SNR, $\mathsf{snr}_i^{-1}$, because the bits over the highest
inverse-SNR are first removed.

Now, let $[r_1,\cdots,r_n]$ be the optimal allocation that minimizes
the inverse system margin $\gamma(R)^{-1}$ for the target bit-rate
$R<R^*$.  Following the algorithm strategy, the optimal allocation for
target bit-rate $R+1$ is obtained adding one bit on subchannel $j$
such that
\begin{align}\label{eq_gup1ann}
  j=\arg\min_i\frac{2^{r_i+1}-1}{\mathsf{snr}_i}\ .
\end{align}
We first prove that
\begin{align}\label{eq_gamr1ann}
  \gamma(R+1)^{-1}=\frac{2^{r_j+1}-1}{\mathsf{snr}_j}\ .
\end{align}
Suppose there exists $j'$ such that
\begin{align}
  \frac{2^{r_{j'}}-1}{\mathsf{snr}_{j'}}>
  \frac{2^{r_j+1}-1}{\mathsf{snr}_j}\ ,
\end{align}
then one bit must be added to subchannel $j$ to obtain $r_j+1$ bits
before adding one bit to subchannel $j'$ to obtain $r_{j'}$ bits which
means that $[r_1,\cdots,r_n]$ is not optimal. As $[r_1,\cdots,r_n]$ is
optimal by definition, it yields
\begin{align}\label{eq_rap1ann}
  \forall i\in[1,n]\quad
  \frac{2^{r_i}-1}{\mathsf{snr}_i}\leq\frac{2^{r_j+1}-1}{\mathsf{snr}_j}
\end{align}
which proves \eqref{eq_gamr1ann}. The first allocated bit is from the
subchannel with the lowest inverse-SNR given by \eqref{eq_gup1ann}
with $r_i=0$ for all $i$.

Comparing \eqref{eq_kup1ann} with \eqref{eq_rap1ann} yields that
$k=j$, and the index subchannel of the first added bit is the same as
the last removed bit. All the intermediate allocations are then
identical with bit-addition and bit-removal methods. There exists only
one way to reach the optimal allocation $R^*$ starting from the empty
loading.

Proof of Lemma \ref{theo_itergap} can also be provided in the
framework of matroid algebraic theory \cite{FASA03, WILS73}.

\section{Proof of Lemma \ref{theo_iter}}\label{ann_iter}

To simplify the notation and without loss of generality the proof is
given with $\beta=1$. Let $[r_1,\cdots,r_n]$ be the optimal allocation
for the target bit-rate $R$ such that $\sum r_i=R$. Let $R+1$ the
new target bit-rate.  We first prove that
$\Delta_i(r_i)=(r_i+1)\mathsf{ber}_i(r_i+1)-r_i\mathsf{ber}_i(r_i)$ is
a good measure at each step of the greedy-type algorithm for the BER
minimization, and finally that $(r_i+1)\mathsf{ber}_i(r_i+1)$ can be
used instead of $\Delta_i(r_i)$.

Starting from the optimal allocation of target bit-rate $R$, the new
target bit-rate $R+1$ is obtained by increasing $r_j$ by one bit
\begin{align}
  \mathsf{ber}(R+1)=\frac{(r_j+1)\mathsf{ber}_j(r_j+1)+\sum\limits_{\substack{i=1\\i\neq j}}^nr_i\mathsf{ber}_i(r_i)}{1+\sum\limits_{i=1}^nr_i}
\end{align}
and, using $\Delta_j$,
\begin{align}
  \mathsf{ber}(R+1)=\frac{\Delta_j(r_j)}{R+1}+\frac{R}{R+1}\mathsf{ber}(R)\,.
\end{align}
The $\mathsf{ber}(R+1)$ which is equal to
$\phi(\{r_i^{(k+1)}\}_{i=1}^n)$ in \eqref{eq_fox} is minimized only if
$\Delta_j(r_j)$ is minimized.  The minimum $\mathsf{ber}(R+1)$ is then
obtained with the increase of one bit in the subchannel $j$ such that
\begin{align}
  j=\arg\min_i\Delta_i(r_i)\,.
\end{align}

To complete the proof by induction, the relation must be true for
$\mathsf{ber}(1)$. This is simply done by recalling that
$\mathsf{ber}_i(0)=0$, and then
\begin{align}
  \min\mathsf{ber}(1)=\min_i\mathsf{ber}_i(1)=\min_i\Delta_i(0)\,.
\end{align}
The convergence of the algorithm to a unique solution needs the
convexity of the function $r_i\mapsto r_i\mathsf{ber}(r_i)$. This
convexity is verified at high SNR.  Appendix~\ref{ann_conv} provides a
more precise domain of validity.

It remains to prove that $(r_i+1)\mathsf{ber}_i(r_i+1)$ can be used
instead of $\Delta_i(r_i)$. In high SNR regime
\begin{align}
  \mathsf{ber}_i(r_i+1)\gg\mathsf{ber}_i(r_i)
\end{align}
and then
\begin{align}
  \lim_{\mathsf{snr}_i\to+\infty}\Delta_i(r_i)=(r_i+1)\mathsf{ber}_i(r_i+1)
\end{align}
which proves the lemma.

In lower SNR regime, the approximation of $\Delta_i$ by
$(r_i+1)\mathsf{ber}_i(r_i+1)$ remains valid and, in practice, the
dissimilarity between allocation using $\Delta_i$ and allocation using
$(r_i+1)\mathsf{ber}_i(r_i+1)$ is null in the domain of validity given
by appendix~\ref{ann_conv}.

\section{Range of convexity of $r_i\mathsf{ber}_i$}\label{ann_conv}

Let
\begin{align}
  f:\ \mathbb{N}&\to\mathbb{R}_+\\
  r_i&\mapsto r_i\mathsf{ber}_i(r_i,\mathsf{snr}_i)\notag
\end{align}
which equals the SER for high SNR regime and Gray mapping. The
function $f$ is a strictly increasing function: $\forall
\mathsf{snr}_i$, $f(r_i)<f(r_i+1)$ because
$\mathsf{ber}(r_i,\mathsf{snr}_i)\leq\mathsf{ber}(r_i+1,\mathsf{snr}_i)$
and $r_i<r_i+1$.  Let $\Delta(r_i)=f(r_i+1)-f(r_i)$, then
\begin{align}
  \Delta(r_i+1)-\Delta(r_i)&=f(r_i+2)-2f(r_i+1)+f(r_i)\\
  &\geq(r_i+1)\big(\mathsf{ber}_i(r_i+2)-2\mathsf{ber}_i(r_i+1)\big)\notag
\end{align}
If $\mathsf{ber}_i(r_i+2)\geq2\mathsf{ber}_i(r_i+1)$ then the function
$f$ is locally convex, or defines a convex hull. This relation is
verified for BER lower than $2\times10^{-2}$ and for all $r_i\geq0$.

\section{Proof of Theorem \ref{theo_cviter}}\label{ann_cviter}

We prove that both metrics used in Lemmas~\ref{theo_itergap} and
\ref{theo_iter} lead to the same subchannel SNR ordering.  Let
\begin{align}
  f(r_i,\mathsf{snr}_i)=\frac{2^{r_i+\beta}-1}{\mathsf{snr}_i}
\end{align}
and
\begin{align}
  g(r_i,\mathsf{snr}_i)=(r_i+\beta)\mathsf{ber}_i(r_i+\beta)\,.
\end{align}
We then have to prove that
\begin{align}\label{eq_ineqann}
  f(r_i,\mathsf{snr}_i)\leq f(r_j,\mathsf{snr}_j)\Leftrightarrow
  g(r_i,\mathsf{snr}_i)\leq g(r_j,\mathsf{snr}_j)\,.
\end{align}
The first inequality yields
\begin{align}
  \frac{\mathsf{snr}_j}{\mathsf{snr}_i}\leq\frac{2^{r_j+\beta}-1}{2^{r_i+\beta}-1}\,.
\end{align}

With square QAM, in high SNR regime and using \eqref{eq_appr}
\begin{align}
  g(r_i,\mathsf{snr}_i)=2\left(1-\frac{1}{\sqrt{2^{r_i+\beta}}}\right)\erfc{\sqrt{\frac{3}{2(2^{r_i+\beta}-1)}\mathsf{snr}_i}}
\end{align}
and it can be approximated by the following valid expression in high
SNR regime
\begin{align}\label{eq_secapproxann}
  g(r_i,\mathsf{snr}_i)=2\erfc{\sqrt{\frac{3}{2(2^{r_i+\beta}-1)}\mathsf{snr}_i}}\,.
\end{align}

The second inequality of \eqref{eq_ineqann} then leads to
\begin{align}
    \frac{\mathsf{snr}_j}{\mathsf{snr}_i}\leq\frac{2^{r_j+\beta}-1}{2^{r_i+\beta}-1}
\end{align}
which is also given by the first inequality. In high SNR regime and
with square QAM, i.e.\ $\beta=2$, $f(\cdot)$ and $g(\cdot)$ lead to
the same subchannel SNR ordering and then
\begin{align}
  \arg\min_if(r_i,\mathsf{snr}_i)=\arg\min_ig(r_i,\mathsf{snr}_i)\,.
\end{align}
This last equation does not hold in low SNR regime (the BER
approximation is not valid) or when the modulations are not square,
i.e.\ when $r_i$ is odd.  Note that \eqref{eq_secapproxann} is not
only a good approximation in high SNR regime, it can also be used with
high modulation orders with moderate SNR regime as it is defined in
Appendix~\ref{ann_conv}.

\section{Proof of Theorem \ref{theo_gap}}\label{ann_gap}

As the infinite norm is not differentiable, we use the $k$ norm with
\begin{align}
  \lim_{k\to+\infty}{\Big(\sum_{i=1}^n\gamma_i^{-k}\Big)}^{\frac{1}{k}}=\max_i(\gamma_i^{-1})\ .
\end{align}
With unconstrained modulations, the Lagrangian of \eqref{eq_pbgap} for
all $k$ is
\begin{align}
  L_k(\{r_i\}_{i=1}^n,\lambda)=\left(\sum_{i=1}^n\frac{{(2^{r_i}-1)}^k}{\mathsf{snr}_i^k}\right)^{\frac{1}{k}}+\lambda\left(R-\sum_{i=1}^nr_i\right).
\end{align}
Let $\lambda'$ such as
\begin{align}
  \lambda'=\left(\sum_{i=1}^n\frac{{(2^{r_i}-1)}^k}{\mathsf{snr}_i^k}\right)^{\frac{k-1}{k}}\frac{\lambda}{\log2}\,.
\end{align}
The optimal condition yields
\begin{align}
  2^{r_i}{(2^{r_i}-1)}^{k-1}=\mathsf{snr}_i^k\lambda'\,.
\end{align}
In asymptotic regime, $r_i\gg1$ and then $2^{r_i}-1\simeq2^{r_i}$. The
equation of the optimal condition can be simplified and
\begin{align}
  r_i=\log_2(\mathsf{snr}_i)+\frac{1}{k}\log_2\lambda'\,.
\end{align}
The Lagrange multiplier is identify using the bit-rate constraint, and
replacing $\lambda'$ in the above equation leads to the solution
\begin{align}
 r_i=\frac{R}{n}+\frac{1}{n}\sum_{j=1}^n\log_2\frac{\mathsf{snr}_i}{\mathsf{snr}_j}\,.
\end{align}
Note that we do not need to calculate the convergence of the solution
with $k\to+\infty$ to obtain the result for the infinite norm. The
result holds for all values of $k$ in asymptotic regime.

With $k=1$, the problem is a sum SNR-gap maximization problem under
peak-power constraint and it can be solved without asymptotic regime
condition.  Note that this sum SNR-gap maximization problem, or sum
inverse SNR-gap minimization problem, under peak-power and bit-rate
constraints is
\begin{align}
  \min\sum_{i=1}^n\gamma_i^{-1}=\min_{\{r_i\}_{i=1}^n}\sum_{i=1}^n(2^{r_i}-1)\frac{\sigma_{W_i}^2}{|h_i|^2Pp_i}
\end{align}
and is very similar to power minimization problem under bit-rate and
SNR-gap constraints exchanging $p_i$ with $\gamma_i^{-1}$
\begin{align}
  \min\sum_{i=1}^np_i=\min_{\{r_i\}_{i=1}^n}\sum_{i=1}^n(2^{r_i}-1)\frac{\sigma_{W_i}^2}{|h_i|^2P\gamma_i^{-1}}\,.
\end{align}
Both problems are identical if $p_i\gamma_i=\alpha$ as it is stated by
Lemma~\ref{theo_eqmarpow}.

\section{Proof of Theorem \ref{theo_lag}}\label{ann_lag}

To prove this theorem, variables $I_i$ and $J_i$ are used instead of
$r_i$ and the bit-rate constraint is
\begin{align}
  R=\sum_{i=1}^n\log_2(I_iJ_i)\,.
\end{align}
With unconstrained QAM, the Lagrangian of \eqref{eq_pbber} is then
\begin{align}
  L(\{I_i,J_i\}_{i=1}^n,\lambda)=&\frac{1}{R}\sum_{i=1}^n\Bigg(2-\frac{1}{I_i}-\frac{1}{J_i}\Bigg)\notag\\&
  \times\erfc{\sqrt{\frac{3}{I_i^2+J_i^2-2}\mathsf{snr}_i}}\notag\\&+
  \lambda\left(R-\sum_{i=1}^n\log_2(I_iJ_i)\right)\,.
\end{align}

Let $X_i\in\{I_i,J_i\}$, then
\begin{align}
  \frac{\partial L}{\partial
    X_i}=X_if(I_i,J_i)+\frac{1}{X_i^2}g(I_i,J_i)-\frac{1}{X_i}\lambda\,,
\end{align}
with
\begin{align}
  f(I_i,J_i)=\frac{1}{R}\Bigg(2-\frac{1}{I_i}-\frac{1}{J_i}\Bigg)\frac{2\sqrt{3\mathsf{snr}_i}\times e^{-\frac{3\mathsf{snr}_i}{I_i^2+J_i^2-2}}}{\sqrt{\pi}(I_i^2+J_i^2-2)^{3/2}}
\end{align}
and
\begin{align}
  g(I_i,J_i)=\frac{1}{R}\erfc{\sqrt{\frac{3\mathsf{snr}_i}{I_i^2+J_i^2-2}}}.
\end{align}

The optimality condition yields
\begin{align}\label{eq_aa}
  \forall i\ (I_i^2-J_i^2)I_iJ_if(I_i,J_i)=(I_i-J_i)g(I_i,J_i)\,.
\end{align}
A trivial solution is $I_i=J_i$ and the other solution must verify
\begin{align}\label{eq_az}
  (I_i+J_i)I_iJ_if(I_i,J_i)-g(I_i,J_i)=0\,.
\end{align}
To find the root of \eqref{eq_az}, let
\begin{align}
  h(x,y)=x\sqrt{y}e^{-y}-\erfc{\sqrt{y}}
\end{align}
with
\begin{align}
  x=\frac{2}{\sqrt{\pi}}\frac{(I_i+J_i)I_iJ_i}{I_i^2+J_i^2-2}\left(2-\frac{1}{I_i}-\frac{1}{J_i}\right)
\end{align}
and
\begin{align}
  y=\frac{3\mathsf{snr}_i}{I_i^2+J_i^2-2}\,.
\end{align}
We will prove that this function is positive in a specific domain.
\begin{enumerate}
\item $\sqrt{y}e^{-y}>\erfc{\sqrt{y}}$ for $y\geq0.334$, then for BER
  lower than $10^{-1}$.
\item $\frac{\sqrt{\pi}}{2}x>1$ for
  $\{I_i,J_i\}\in[1,+\infty)^2$ and $I_i\neq1$ or $J_i\neq1$,
  and $\lim\limits_{I_i,J_i\to1}\frac{\sqrt{\pi}}{2}x=1^+$.
\end{enumerate}
Then, in the domain defined by
\begin{align}\label{eq_ae}
  \{I_i,J_i\}\in[1,+\infty)^2\wedge\mathsf{ber}_i\leq0.1
\end{align}
$h(x,y)$ is positive and \eqref{eq_az} has no solution. Thus, the only
one solution of \eqref{eq_aa} with \eqref{eq_ae} is $I_i=J_i$. As we
will see later the domain of \eqref{eq_ae} is less restrictive than
the asymptotic one.

The problem is now to allocate bits with square QAM. The following
upper bound is used
\begin{align}
  \mathsf{ber}(r_i)=\frac{2}{r_i}\erfc{\sqrt{\frac{3\mathsf{snr}_i}{2(2^{r_i}-1)}}}.
\end{align}
Note that this upper bound is a tight approximation with high SNR and
with high modulation orders. The Lagrangian is
\begin{align}
  L(\{r_i\}_{i=1}^n,\lambda)&=\frac{2}{R}\sum_{i=1}^n\erfc{\sqrt{\frac{3\mathsf{snr}_i}{2(2^{r_i}-1)}}}\notag\\&\quad+\lambda\left(R-\sum_{i=1}^kr_i\right).
\end{align}
and its derivative is
\begin{align}
  \frac{\partial L}{\partial r_i}=\frac{\ln2}{\sqrt{\pi}}\frac{2^{r_i}}{2^{r_i}-1}\sqrt{\frac{3\mathsf{snr}_i}{2(2^{r_i}-1)}}e^{-\frac{3\mathsf{snr}_i}{2(2^{r_i}-1)}}-\lambda\,.
\end{align}
Let $\forall i$ $r_i\gg1$, then $2^{r_i}-1\simeq2^{r_i}$ and the
optimality condition yields
\begin{align}
  -\frac{3\mathsf{snr}_i}{2^{r_i}}e^{-\frac{3\mathsf{snr}_i}{2^{r_i}}}=-\frac{2\lambda^2\pi}{\ln^22}\,.
\end{align}
With reliable communication over the subchannel $i$, the Shannon's
relation states that $r_i\leq\log_2(1+\mathsf{snr}_i)$ and
$\frac{3\mathsf{snr}_i}{2^{r_i}}\geq\frac{3}{2}$ because $r_i\geq1$.
The relation between $r_i$ and $\lambda$ is then bijective and the
real branch $W_{-1}$ of the Lambert function \cite{CORL96} can be used
with no possibility for confusion
\begin{align}\label{eq_ar}
  r_i=\log_2(3\mathsf{snr}_i)-\log_2\left(-W_{-1}\left(-\frac{2\lambda^2\pi}{\ln^22}\right)\right)\,.
\end{align}
With the bit-rate constraint $R=\sum\limits_ir_i$, we can write
\begin{align}
  -\log_2\left(-W_{-1}\left(-\frac{2\lambda^2\pi}{\ln^22}\right)\right)=\frac{R}{n}-\frac{1}{n}\sum_{i=1}^n\log_2(3\mathsf{snr}_i)
\end{align}
and with \eqref{eq_ar}
\begin{align}
  r_i=\frac{R}{n}+\frac{1}{n}\sum_{j=1}^n\log_2\frac{\mathsf{snr}_i}{\mathsf{snr}_j}\,.
\end{align}
This result is obtained with square QAM in asymptotic regime (high
modulation orders and high SNR) which is a more restrictive domain
than that of \eqref{eq_ae}.

%%% Local Variables: 
%%% mode: latex
%%% TeX-master: "article"
%%% End: